\documentstyle[rotating,epic,eepic,aps,epsfig,pre,twocolumn,amstex]{revtex}
  \title{Competition of mixing  and segregation in rotating cylinders}
  \author{Christian M. Dury and Gerald H. Ristow}
  \address{Fachbereich Physik, Philipps-Universit\"at, Renthof 6, 
           35032 Marburg, Germany}
\date{received March 5, 1998; revised May 28, 1998}

\draft
  \newcommand{\tenrm}{\normalsize}
  \newcommand{\grdraft}{true}
  \ifx\grdraft\undefined
  \else
  \fi
\begin{document}
\maketitle
\begin{abstract}
Using discrete element methods, we study numerically the dynamics of the size 
segregation process of binary particle mixtures in three-dimensional rotating
drums,  operated in the continuous flow regime. Particle rotations are included
and we focus on different volume filling fractions of the drum to  study the
interplay between the competing phenomena of mixing and segregation. It is
found that segregation is best for a more than half-filled drum due to the
non-zero width of  the fluidized layer.  For different particle size ratios, it
is found that radial segregation occurs for any arbitrary small particle size
difference and  the final amount of segregation shows a linear dependence on
the size ratio of the two particle species. To quantify the interplay between
segregation and mixing, we investigate the dynamics of the center of mass
positions for each particle component. Starting with initially separated
particle groups  we find that no mixing of the component is necessary in order
to obtain a radially segregated core.
\end{abstract}

\section{Introduction}
When granular materials are placed in rotating cylinders, different flow
dynamics are  observed. The major portion of the particles undergoes a solid
body rotation by following the cylinder motion. Close to the free surface, a
downhill particle flow is observed where the flow dynamics depends on the
rotation speed of the  cylinder~\cite{nityanand86,rajchenbach90}. For low
rotation speeds, the surface flow consists of individual avalanches called
discrete avalanche regime. With increasing rotation speed the separation time
of avalanches will decrease until no individual avalanches are detectable. A
nearly constant particle flow is found along the free surface and this regime
is consequently termed continuous flow regime. Close to this transition, the
surface can be well approximated by a straight plane, which can be used to
determine the surface angle. For even higher rotation speeds,  cascading,
cataracting and centrifuging particle motion is also
observed~\cite{nityanand86}.  Experiments  performed in the discrete avalanche
regime  using mono-disperse particles show that  the mixing time depends
strongly on the volume filling fraction of the cylinder and no mixing is seen
for an exactly half-filled cylinder~\cite{metcalfe95}. A theoretical
description could be given which is based on the mixing between 
wedges~\cite{metcalfe95,peratt96,dorogovtsev98}. On the other hand, when a
mixture of particles which differ in size or density are placed in a rotating
cylinder, the denser or smaller particles will concentrate in a central region
close to the free surface after only a few rotations, which is termed {\em
radial segregation}. This was studied experimentally and numerically for
varying size ratios~\cite{clement95,cantelaube95,baumann95,hill97,dury97} and
density ratios~\cite{ristow94,metcalfe96,khakhar97b}. The amount  and direction
of segregation depends on the rotation rate~\cite{nityanand86}.  However,
radial segregation is always observed in the continuous flow regime, regardless
of  the filling fraction of the cylinder and we are investigating numerically
the interplay of the two competing phenomena of mixing and segregation, a
problem first studied by  Rose~\cite{rose59} and restated by
Behringer~\cite{behringer95}. We are particularly interested in the dependence
of the total amount of segregation on the size ratio of glass beads and how the
segregation behaves for size ratios close to one.

The paper is organized in the following way: After a brief motivation of the
physical system in mind in the beginning of the next section, we will explain
our numerical model and the physical meaning of the parameters in remainder of
section~\ref{sdrei}.  In order to model  the dynamics of glass beads correctly
with our model, we have to include rolling friction. The details of our
implementation are given in the appendix. We define a  order parameter which
allows to  quantify the amount and the speed of radial segregation. The
dynamics of this order parameter are discussed in detail in section~\ref{svier}
for different particle size ratios and as function of the volume filling
fraction of the cylinder. In addition, we start with an initial configuration
similar to the one used in the original experiment on the mixing of
mono-disperse particles~\cite{metcalfe95} to demonstrate how the radial
segregation competes with the mixing process.

\section{Motivation and Numerical Model}\label{sdrei}
In many radial and axial segregation experiments, one of the components are
glass beads. They are commercially available in  large quantities and can
easily be sieved to nearly uniform size distribution. When sufficiently large,
the cohesion forces are negligible. In the continuous flow regime the dynamic
angle of repose for glass spheres is independent of the rotation speed of the
cylinder as long as the surface remains flat~\cite{zik94,dury97d}. Since they
are nearly perfectly round, particle rotations are an important degree of
freedom and cannot be neglected in a theoretical or numerical description.
We use three-dimensional discrete element methods, also known as {\em granular
dynamics}, this gives us the advantage to vary particle size ratios freely,
whereas in experiments only a limited number of particle diameters and
densities are available.

We want to investigate radial segregation, therefore we use only a cylinder
with periodic  boundary conditions along the rotational axis.  In order to
determine the minimal length of this cylinder, to get no artifacts due to the
periodic boundary conditions,  we have investigated the range of the boundary
effects in a previous study~\cite{dury97d}.

\subsection{Forces during Collisions}
Each particle $i$ is approximated by a sphere with radius $R_i$. Only  contact
forces during collisions are considered and the particles are allowed to
rotate; we also include rolling resistance to our model to correctly  describe
the dynamics of glass beads (see appendix). The forces acting on particle $i$
during a collision with particle $j$ are 
\begin{equation}
  F_{ij}^n = - \tilde{Y} \ (R_i + R_j - \vec{r}_{ij}\hat{n}) -
               \gamma_n \vec{v}_{ij} \hat{n}
  \label{eq: fn}
\end{equation}
in the normal direction ($\hat{n}$) and
\begin{equation}
  F_{ij}^s = -\min(\gamma_s \vec v_{ij}\cdot\hat{s}(t), \mu|F_{ij}^n|) \ .
  \label{fric1}
\end{equation}
in the tangential direction ($\hat{s}$) of shearing.  In Eq.~(\ref{eq: fn})
$\gamma_n$ represent the dynamic damping coefficient  and Eq.~(\ref{fric1})
$\gamma_s$ represent the dynamic friction force in the tangential direction.
$\vec{r}_{ij}$ represents the vector  joining both centers of mass,
$\vec{v}_{ij}$  represents the relative motion of  the two particles, and
$\tilde{Y}$ is related to the Young Modulus of the  investigated material.
Dynamic friction in this model is defined to be proportional to the relative
velocity of the particles in the tangential direction.

During particle--wall contacts, the wall is treated as a particle with infinite
mass and radius. In the normal direction, Eq.~(\ref{eq: fn}) is applied,
whereas in the tangential direction, the static friction force
\begin{equation}
  \tilde{F}_{ij}^s = -\min(k_s \int \vec v_{ij}\cdot\hat{s}(t) dt,
  \mu|F_{ij}^n|)
  \label{fric2}
\end{equation}
is used. This is motivated by the observation that when particles flow along
the free surface, they dissipate most of their energy in collisions and can
come to rest in voids left by other particles. This is not possible at the 
cylinder walls. 
In order to avoid additional artificial particles at the walls
we use a static friction law to avoid slipping and allowing
for a static surface angle when the rotation is stopped. 
Both tangential forces
are limited by the Coulomb criterion, see Eqs.~(\ref{fric1}) and 
(\ref{fric2}), which states that the magnitude of the tangential force cannot
exceed the magnitude of the normal force multiplied by the friction coefficient
$\mu$.  Even though the experimentally measured friction coefficient for fresh
glass beads is  $\mu=0.092$~\cite{foerster94} this can only be viewed as a
lower bound in our case due to the wear of material caused by the uncountable
bead collisions in the course of the experiment. For particle--particle
collisions we use $\mu=0.19$, and for particle--wall collisions,
\mbox{$\mu_w=0.6$}.  The coefficient of restitution for wall collisions is set
to 0.97 and to 0.831 for particle--particle collisions, which are the measured
values for glass beads~\cite{foerster94} and the density was set to
$\rho=2.5\frac{g}{cm^3}$. In order to  save computer time, we set $\tilde{Y}$
to \mbox{$6\cdot 10^4$ P$\!$a~m} which is about one order of magnitude softer
than glass, but  we checked that this has no effect on the investigated 
properties of the
material~\cite{dury97c}. This gives a contact time during collisions of
$1.1\cdot 10^{-6}$ s, which is still quite small.
The total number of particles we used were up to 17000.

\subsection{Order parameter}
To compare the quality and speed of  the segregation of different runs with
different  parameters it is desirable to have only one order parameter to
investigate all of these~\cite{dury97}.  For this we divide the cylinder with
Diameter $D$ into $n$ concentric hollow cylinders with  thickness
$\Delta=\frac{D}{2n}$ and measure the number density of the smaller particles, 
$\rho_i$, in each hollow cylinder $i=1,\ldots,n$ (see Fig.~\ref{cylinder}). We
choose as such an order parameter $q$ the sum over all positive deviations of
$\rho_i$ above the  mean number density $\rho_0={(\text{Number of small
particles})}/{(f (D/2)^2 \pi)}$ in a cylinder with a filling fraction $f$,
normalized with respect to the ideally segregated case,  where all inner
cylinders $i = 1,\ldots,c $ are composed only out of small particles shown in
grey in Fig.~\ref{cylinder} and in the outer cylinders $i = c+1,\ldots,n$  only
out of large particles. To get reasonable results, we  take $\Delta$ to be of
the order of the diameter of the larger particles,  which gives in our case
$n=11$. For smaller values of $n$ the resolution gets worse and for higher $n$
fluctuations  become more pronounced, since there would be ``empty'' shells
with no particle centers in it. For  varying  $n$ around 11, $q$ has an 
over-all error of 0.03 for $n\in[9,20]$.\\
%
\ifx\grdraft\undefined  \else
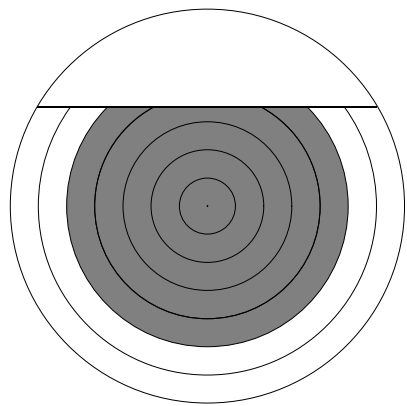
\begin{figure}[htb]
\centering
\setlength{\unitlength}{1.5mm}
\begin{picture}(35,37)(-17.5,-18.5)
\put(0,0){\circle{35}}
\put(0,0){\arc{30}{-0.623}{3.764}}
\put(0,0){\shade\arc{25}{-0.775}{3.917}}
\put(0,0){\arc{20}{-1.065}{4.207}}
\put(0,0){\arc{20}{-1.065}{4.207}}
\put(0,0){\circle{15}}
\put(0,0){\circle{10}}
\put(0,0){\circle{5}}
\put(0,0){\circle*{0.1}}
\put(-15,8.75){\line(1,0){30}}
\end{picture}
\caption{Cross section through a more than half--filled cylinder.}
\label{cylinder}
\end{figure}
\fi
To illustrate this procedure, we show in Fig.~\ref{density} (where we used
eleven concentric shells)  the packing fraction of small ($\Diamond$) and large
($+$) particles in each cylinder for a volume filling fraction of $66.4\%$. 
The volume filling fraction is defined as the the ratio of the volume occupied 
%
\ifx\grdraft\undefined  \else
\begin{figure}[t]
\noindent	
{\centering 
\begin{picture}(1500,900)(0,50)
\tenrm
\thicklines \path(199,179)(219,179)
\thicklines \path(954,179)(934,179)
\put(177,179){\makebox(0,0)[r]{0}}
\thicklines \path(199,288)(219,288)
\thicklines \path(954,288)(934,288)
\put(177,288){\makebox(0,0)[r]{0.1}}
\thicklines \path(199,398)(219,398)
\thicklines \path(954,398)(934,398)
\put(177,398){\makebox(0,0)[r]{0.2}}
\thicklines \path(199,507)(219,507)
\thicklines \path(954,507)(934,507)
\put(177,507){\makebox(0,0)[r]{0.3}}
\thicklines \path(199,617)(219,617)
\thicklines \path(954,617)(934,617)
\put(177,617){\makebox(0,0)[r]{0.4}}
\thicklines \path(199,726)(219,726)
\thicklines \path(954,726)(934,726)
\put(177,726){\makebox(0,0)[r]{0.5}}
\thicklines \path(199,836)(219,836)
\thicklines \path(954,836)(934,836)
\put(177,836){\makebox(0,0)[r]{0.6}}
\thicklines \path(199,945)(219,945)
\thicklines \path(954,945)(934,945)
\put(177,945){\makebox(0,0)[r]{0.7}}
\thicklines \path(199,179)(199,199)
\thicklines \path(199,945)(199,925)
\put(199,134){\makebox(0,0){0}}
\thicklines \path(307,179)(307,199)
\thicklines \path(307,945)(307,925)
\put(307,134){\makebox(0,0){0.5}}
\thicklines \path(415,179)(415,199)
\thicklines \path(415,945)(415,925)
\put(415,134){\makebox(0,0){1}}
\thicklines \path(523,179)(523,199)
\thicklines \path(523,945)(523,925)
\put(523,134){\makebox(0,0){1.5}}
\thicklines \path(630,179)(630,199)
\thicklines \path(630,945)(630,925)
\put(630,134){\makebox(0,0){2}}
\thicklines \path(738,179)(738,199)
\thicklines \path(738,945)(738,925)
\put(738,134){\makebox(0,0){2.5}}
\thicklines \path(846,179)(846,199)
\thicklines \path(846,945)(846,925)
\put(846,134){\makebox(0,0){3}}
\thicklines \path(954,179)(954,199)
\thicklines \path(954,945)(954,925)
\put(954,134){\makebox(0,0){3.5}}
\thicklines \path(199,179)(954,179)(954,945)(199,945)(199,179)
\put(45,562){\makebox(0,0)[l]{\shortstack{$\eta$}}}
\put(576,67){\makebox(0,0){distance[cm]}}
\put(780,312){\makebox(0,0)[r]{small}}
\thinlines \path(802,312)(910,312)
\thinlines \path(230,768)(230,768)(293,743)(356,676)(419,668)(482,683)(545,671)(608,642)(671,624)(734,554)(797,451)(860,345)
\put(230,768){\raisebox{-1.2pt}{\makebox(0,0){$\Diamond$}}}
\put(293,743){\raisebox{-1.2pt}{\makebox(0,0){$\Diamond$}}}
\put(356,676){\raisebox{-1.2pt}{\makebox(0,0){$\Diamond$}}}
\put(419,668){\raisebox{-1.2pt}{\makebox(0,0){$\Diamond$}}}
\put(482,683){\raisebox{-1.2pt}{\makebox(0,0){$\Diamond$}}}
\put(545,671){\raisebox{-1.2pt}{\makebox(0,0){$\Diamond$}}}
\put(608,642){\raisebox{-1.2pt}{\makebox(0,0){$\Diamond$}}}
\put(671,624){\raisebox{-1.2pt}{\makebox(0,0){$\Diamond$}}}
\put(734,554){\raisebox{-1.2pt}{\makebox(0,0){$\Diamond$}}}
\put(797,451){\raisebox{-1.2pt}{\makebox(0,0){$\Diamond$}}}
\put(860,345){\raisebox{-1.2pt}{\makebox(0,0){$\Diamond$}}}
\put(856,312){\raisebox{-1.2pt}{\makebox(0,0){$\Diamond$}}}
\put(780,267){\makebox(0,0)[r]{large}}
\thicklines \path(802,267)(910,267)
\thicklines \path(230,233)(230,233)(293,272)(356,341)(419,400)(482,365)(545,374)(608,412)(671,424)(734,502)(797,610)(860,712)
\put(230,233){\makebox(0,0){$+$}}
\put(293,272){\makebox(0,0){$+$}}
\put(356,341){\makebox(0,0){$+$}}
\put(419,400){\makebox(0,0){$+$}}
\put(482,365){\makebox(0,0){$+$}}
\put(545,374){\makebox(0,0){$+$}}
\put(608,412){\makebox(0,0){$+$}}
\put(671,424){\makebox(0,0){$+$}}
\put(734,502){\makebox(0,0){$+$}}
\put(797,610){\makebox(0,0){$+$}}
\put(860,712){\makebox(0,0){$+$}}
\put(856,267){\makebox(0,0){$+$}}
\put(780,222){\makebox(0,0)[r]{total}}
\Thicklines \path(802,222)(910,222)
\Thicklines \path(230,822)(230,822)(293,836)(356,838)(419,889)(482,869)(545,866)(608,875)(671,869)(734,877)(797,882)(860,878)
\put(230,822){\raisebox{-1.2pt}{\makebox(0,0){$\Box$}}}
\put(293,836){\raisebox{-1.2pt}{\makebox(0,0){$\Box$}}}
\put(356,838){\raisebox{-1.2pt}{\makebox(0,0){$\Box$}}}
\put(419,889){\raisebox{-1.2pt}{\makebox(0,0){$\Box$}}}
\put(482,869){\raisebox{-1.2pt}{\makebox(0,0){$\Box$}}}
\put(545,866){\raisebox{-1.2pt}{\makebox(0,0){$\Box$}}}
\put(608,875){\raisebox{-1.2pt}{\makebox(0,0){$\Box$}}}
\put(671,869){\raisebox{-1.2pt}{\makebox(0,0){$\Box$}}}
\put(734,877){\raisebox{-1.2pt}{\makebox(0,0){$\Box$}}}
\put(797,882){\raisebox{-1.2pt}{\makebox(0,0){$\Box$}}}
\put(860,878){\raisebox{-1.2pt}{\makebox(0,0){$\Box$}}}
\put(856,222){\raisebox{-1.2pt}{\makebox(0,0){$\Box$}}}
\end{picture} }
 \caption{Volume fraction of small and large particles in each cylinder for
a volume filling fraction of 66.4\%.}
\label{density}
\end{figure}
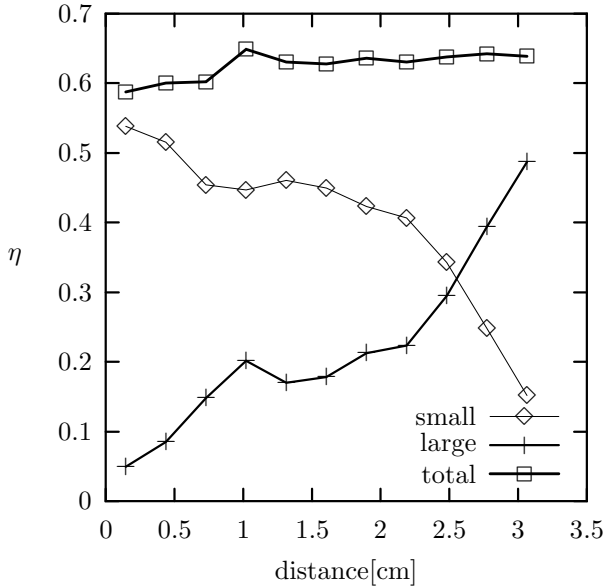
\fi
\noindent
by the granular material to the cylinder volume, reading
$f=\frac{V_{\text{occupied}}}{\left(\frac{D}{2}\right)^2\pi L}$, where the 
occupied volume depends on the packing fraction $\eta$ via 
$V_{\text{occupied}}=\frac{1}{\eta}\sum_{\text{all
particles}}\frac{4}{3}r^3\pi$. Hence the packing fraction $\eta$ is determined 
by 
\[ \eta = \frac{\sum_{\text{all particles}}\frac{4}{3}r^3\pi}{f\cdot 
   (\frac{D}{2})^2\pi L} \qquad ;\]
i.e. filling the drum with a certain filling fraction and summing up the
volumes  of each particle. We found that this packing fraction $\eta$  does not
change with the amount of achieved segregation. For Fig.~\ref{density} the
particle size ratio of small and large particles $\Phi:=\frac{r}{R}$ was
$\Phi=\frac{0.75\textrm{mm}}{1.5\textrm{mm}}=0.5$  and the cylinder was rotated
for $2\frac{1}{2}$ revolutions which gives a nearly complete radial
segregation. The small particles,  denoted by $\Diamond$, are mostly found in
the middle of the cylinder, whereas the large particles, denoted by $+$, show a
higher concentration in the outer cylinders. Also shown is the total volume
occupied by all spheres,  denoted by $\Box$, which gives an average value in
each cylinders of $\eta=0.64$ which is close to the value for a random particle
packing. \section{Size Segregation}\label{svier} For our simulation we are
using a cylinder with diameter $D=7$cm and periodic boundary conditions along
the rotation axis. The cylinder length was $2.5$cm which was sufficiently long
to ensure that boundary effects are negligible~\cite{dury97d}. The cylinder was
filled with a binary mixture of large beads having a radius of $R=1.5$mm and 
small beads $r\in \{0.75$mm$,1.0$mm$,1.25$mm$\}$, where the small particles can
have a concentration of 50\% or 33\%  by volume.  The aspect ratio of the drum
diameter $D$ and the average particle diameter $2r$ is $D/(2r)=28$, this is of
the order of laboratory experiments, where we have $D/(2r)=25$ up to  $40$ for
the example in Ref.~\cite{nityanand86}. The density of our beads is always
$\rho=2.5\frac{g}{cm^3}$. For our value of $\Omega=15$rpm, we are in the
continuous flow regime with a flat free surface and the Froude number is
\[ Fr=\frac{\langle v\rangle^2}{lg}=(2.3\ldots 9.1) \times 10^{-2} \qquad,\]
so we can neglect inertia effects.

%
\ifx\grdraft\undefined  \else
\begin{figure}[t]
\noindent	
{\centering 
\begin{picture}(1500,900)(0,50)
\tenrm
\thicklines \path(199,179)(219,179)
\thicklines \path(954,179)(934,179)
\put(177,179){\makebox(0,0)[r]{0}}
\thicklines \path(199,332)(219,332)
\thicklines \path(954,332)(934,332)
\put(177,332){\makebox(0,0)[r]{0.2}}
\thicklines \path(199,485)(219,485)
\thicklines \path(954,485)(934,485)
\put(177,485){\makebox(0,0)[r]{0.4}}
\thicklines \path(199,639)(219,639)
\thicklines \path(954,639)(934,639)
\put(177,639){\makebox(0,0)[r]{0.6}}
\thicklines \path(199,792)(219,792)
\thicklines \path(954,792)(934,792)
\put(177,792){\makebox(0,0)[r]{0.8}}
\thicklines \path(199,945)(219,945)
\thicklines \path(954,945)(934,945)
\put(177,945){\makebox(0,0)[r]{1}}
\thicklines \path(199,179)(199,199)
\thicklines \path(199,945)(199,925)
\put(199,134){\makebox(0,0){0}}
\thicklines \path(307,179)(307,199)
\thicklines \path(307,945)(307,925)
\put(307,134){\makebox(0,0){5}}
\thicklines \path(415,179)(415,199)
\thicklines \path(415,945)(415,925)
\put(415,134){\makebox(0,0){10}}
\thicklines \path(523,179)(523,199)
\thicklines \path(523,945)(523,925)
\put(523,134){\makebox(0,0){15}}
\thicklines \path(630,179)(630,199)
\thicklines \path(630,945)(630,925)
\put(630,134){\makebox(0,0){20}}
\thicklines \path(738,179)(738,199)
\thicklines \path(738,945)(738,925)
\put(738,134){\makebox(0,0){25}}
\thicklines \path(846,179)(846,199)
\thicklines \path(846,945)(846,925)
\put(846,134){\makebox(0,0){30}}
\thicklines \path(954,179)(954,199)
\thicklines \path(954,945)(954,925)
\put(954,134){\makebox(0,0){35}}
\thicklines \path(199,179)(954,179)(954,945)(199,945)(199,179)
\put(45,562){\makebox(0,0)[l]{\shortstack{q(t)}}}
\put(576,67){\makebox(0,0){t[s]}}
\put(485,903){\makebox(0,0)[r]{50\% filled}}
\thinlines \path(507,903)(615,903)
\thinlines \path(231,252)(231,252)(235,252)(240,262)(240,273)(244,254)(249,233)(251,268)(256,286)(260,242)(262,253)(267,313)(271,296)(273,338)(278,394)(282,426)(284,451)(289,497)(293,426)(295,386)(300,381)(304,404)(306,491)(311,381)(315,374)(318,385)(322,423)(326,403)(329,475)(333,495)(338,603)(340,649)(345,521)(349,490)(351,512)(356,546)(360,518)(363,523)(367,475)(372,451)(374,503)(378,463)(383,517)(385,584)(390,580)(394,688)(396,692)(401,618)(405,563)(408,572)(412,625)
\thinlines \path(412,625)(416,534)(419,558)(423,506)(426,545)(431,551)(435,570)(437,545)(442,574)(446,694)(448,718)(453,705)(457,602)(460,591)(464,620)(468,585)(471,596)(475,571)(479,554)(482,564)(486,562)(490,594)(493,622)(497,626)(501,716)(504,771)(508,755)(512,692)(515,663)(519,619)(524,627)(528,597)(530,613)(535,597)(539,597)(541,631)(545,691)(550,664)(552,719)(556,665)(561,752)(563,694)(567,738)(572,679)(574,674)(578,690)(583,642)(585,629)(589,613)(594,629)(596,620)
\thinlines \path(596,620)(600,674)(605,682)(607,709)(611,734)(616,772)(618,766)(622,711)(627,683)(629,742)(633,655)(638,645)(640,628)(644,642)(649,651)(651,679)(655,688)(660,742)(662,762)(666,710)(671,745)(673,739)(677,686)(682,724)(684,760)(688,678)(693,640)(695,633)(699,615)(704,643)(706,694)(710,746)(715,776)(717,783)(722,753)(726,696)(728,684)(733,698)(737,663)(739,731)(744,704)(748,640)(750,663)(755,648)(759,684)(761,682)(766,705)(770,805)(771,782)(776,746)(780,712)
\thinlines \path(780,712)(782,714)(787,729)(791,689)(795,673)(798,676)(802,680)(806,662)(809,653)(813,678)(817,662)(820,687)(824,734)(828,732)(831,811)(835,701)(840,685)(844,739)(846,712)(851,705)(855,701)(857,703)(862,680)(866,680)(868,662)(873,687)(877,751)(880,774)(884,764)(888,719)(891,686)(895,710)(900,656)(902,642)(906,661)(911,721)(913,713)(917,688)(922,665)(924,720)(928,716)(933,777)(935,743)(939,739)(944,693)(946,693)(950,717)(954,676)
\put(485,858){\makebox(0,0)[r]{80\% filled}}
\thicklines \path(507,858)(615,858)
\thicklines \path(231,252)(231,252)(235,252)(236,252)(241,254)(242,254)(246,250)(248,249)(252,250)(253,249)(258,258)(259,254)(263,260)(265,258)(269,251)(271,252)(275,264)(276,259)(281,267)(282,273)(287,286)(288,284)(293,306)(294,310)(298,311)(300,318)(304,339)(306,339)(310,336)(311,331)(316,323)(317,325)(322,336)(323,334)(327,344)(331,343)(333,343)(337,342)(339,344)(343,353)(344,353)(349,362)(350,364)(355,375)(356,378)(360,369)(361,379)(366,383)(367,382)(372,405)(373,408)
\thicklines \path(373,408)(377,422)(378,419)(383,406)(384,403)(388,402)(389,406)(394,406)(395,406)(400,404)(401,407)(405,414)(406,419)(411,420)(412,417)(416,421)(418,420)(422,437)(423,440)(428,432)(429,432)(433,440)(434,440)(439,442)(440,449)(444,474)(446,475)(450,470)(451,471)(456,456)(457,454)(461,459)(462,461)(467,459)(468,455)(473,459)(474,459)(478,468)(479,476)(484,477)(485,473)(489,480)(490,481)(495,493)(496,493)(501,490)(502,490)(506,497)(506,497)(511,511)(512,516)
\thicklines \path(512,516)(516,515)(517,516)(522,510)(523,505)(528,504)(529,505)(533,512)(534,508)(539,506)(540,507)(544,520)(545,520)(550,524)(551,521)(556,518)(557,519)(561,526)(562,527)(567,536)(568,537)(572,526)(573,525)(578,540)(579,539)(584,553)(585,550)(589,543)(590,545)(595,540)(596,537)(600,540)(601,542)(606,546)(607,541)
\thinlines \path(239,179)(245,205)(252,236)(260,266)(268,294)(275,320)(283,344)(291,367)(298,389)(306,409)(313,428)(321,445)(329,462)(336,477)(344,492)(352,505)(359,518)(367,530)(374,541)(382,552)(390,561)(397,571)(405,579)(413,587)(420,595)(428,602)(435,609)(443,615)(451,621)(458,626)(466,631)(474,636)(481,641)(489,645)(496,649)(504,652)(512,656)(519,659)(527,662)(535,665)(542,668)(550,670)(557,673)(565,675)(573,677)(580,679)(588,681)(596,682)(603,684)(611,685)(618,687)
\thinlines \path(618,687)(626,688)(634,689)(641,690)(649,692)(657,693)(664,693)(672,694)(679,695)(687,696)(695,697)(702,697)(710,698)(718,699)(725,699)(733,700)(740,700)(748,701)(756,701)(763,702)(771,702)(779,702)(786,703)(794,703)(801,703)(809,703)(817,704)(824,704)(832,704)(840,704)(847,705)(855,705)(862,705)(870,705)(878,705)(885,705)(893,706)(901,706)(908,706)(916,706)(923,706)(931,706)(939,706)(946,706)(954,706)
\end{picture} }
\caption{Typical time series of the order parameter $q$ for 
two different filling fractions of the cylinder
with $1.0$mm and $1.5$mm beads.}
\label{q_time_series}
\end{figure}
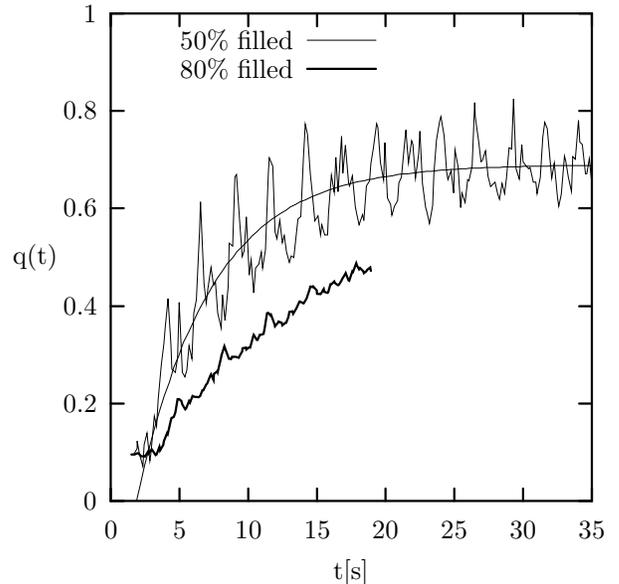
\fi
\subsection{Time evolution of the order parameter}\label{4a}
Usually the initial state is a random mixture of small and large particles
which gives a value of  $q\approx 0$.  For our value of $\Omega=15$rpm,  we are
in the continuous flow regime with a flat free surface and the order parameter
will show a global trend of increasing in time and saturates on the  long run
when the cylinder rotation is started. A typical time evolution of $q$ using a
50\% volume fraction of 1.0mm smaller particles is shown in
Fig.~\ref{q_time_series}. The general trend can be well approximated by an
exponential saturating function of the form
\begin{equation}
q(t)=q_\infty ( 1- e^{-t/t_c})
\label{fit} \end{equation}
with a characteristic segregation time $t_c$ and a final amount of segregation
$q_\infty$. The best fit to the data points for a  50\%  filling of each of
$1.0$mm and $1.5$mm beads was obtained for the parameters $t_c=(6.1\pm0.3)$s
and $q_\infty =0.644\pm0.040$ which was added in Fig.~\ref{q_time_series}.

When working in the discrete avalanche regime, it was found in
experiments~\cite{metcalfe95}  and mathematical
models~\cite{peratt96,dorogovtsev98} that the least amount of geometrical
mixing is given for a half-filled drum which could be explained by the
avalanche mixing of wedges. Peratt and Yorke~\cite{peratt96}  applied their
model also to the continuous avalanche regime by taking the limit of an
infinitely thin  flowing layer and no change in the angle of repose during the
flow. This should lead to ``infinitely'' many avalanches during one revolution
and with a finite avalanche duration all these avalanches are superposed and
could lead to continuous avalanches (steady flow). Motivated by these findings,
we expected for our setup to get the least geometrical mixing for filling
fractions around 50\% as well and the strong  fluctuations which are most
pronounced for a filling fraction of 50\% are a clear sign for  slow
geometrical mixing.
%
\ifx\grdraft\undefined  \else
\begin{figure}[t]
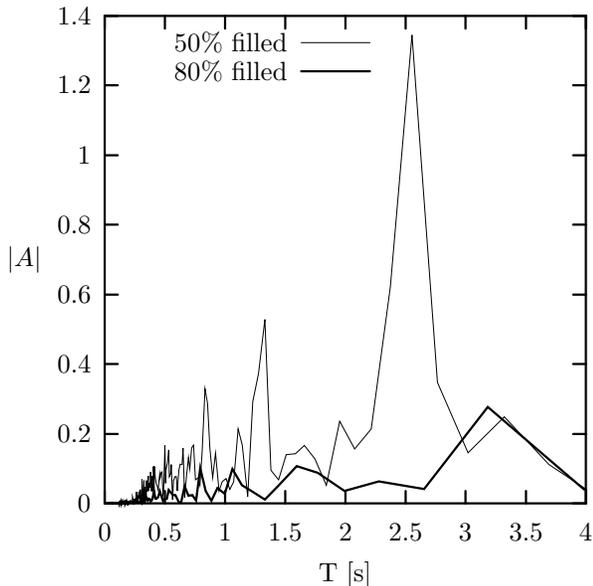

\noindent	
{\centering \input fourier.tex }
\caption{Fourier transform of $q(t)$ of Fig.~\ref{q_time_series} .}
\label{q_fourier}
\end{figure}
\fi

Fourier-transforming these fluctuations gives  main peaks for the 50\% and the
80\% filled drum at $T=(2.56\pm 0.09)s$ and $T=(3.18\pm 0.15)s$, albeit the
peak for the 80\% filled case is much less pronounced than in the 50\%  case as
is shown in Fig.~\ref{q_fourier}; higher harmonics are also visible. These times
are exactly the time it takes for a particle to make one revolution, i.e. to
appear at the same spot again, given as
\begin{equation} 
T=\frac{\Omega}{\alpha}+\frac{l}{\langle v\rangle}
\label{period}
\end{equation}
where $\Omega$ is the angular velocity of the drum, $\alpha$ the  arc where the
particle is in the solid block, $l$ the length of the fluidized layer  and
$\langle v\rangle$ the average velocity of the particles in the fluidized
layer.  For a half-filled drum this gives $\alpha =\pi$ and l=$2R$ . From this
we calculate $\langle v\rangle$ to be $\langle v\rangle=12.5\frac{cm}{s}$ in
each case. The depth of the fluidized layer is determined as in
Ref.\cite{nakagawa93,ristow96} by looking at  the velocity profile along  a
line through the center of the drum and perpendicular to the free surface.  The
depth of the fluidized layer is the distance from the free surface to the point
where the velocity  profile reaches its zero value.

In Fig.~\ref{q_time_series}, the thin line corresponds to a half-filled
cylinder and due to a slight asymmetric start configuration, which persists due
to the bad geometric mixing, large fluctuations which decrease in time are
visible. For other filling ratios of the  cylinder the asymmetry in the
beginning gets erased by the geometrical mixing and the fluctuations are
suppressed or are decaying in a rapid way. This is illustrated by the thick
line in Fig.~\ref{q_time_series} which is for a  volume filling fraction of
80\% and as expected the fluctuations have a much smaller amplitude and do not
show such a pronounced  periodicity.

\subsection{Dependence on the filling fraction}
We will now turn to the pre-factor of the exponential fit in Eq.~(\ref{fit}),
$q_\infty$, which quantifies the final amount of segregation and study its
dependence on the volume filling fraction and on the particle size ratio. This
is shown in Fig.~\ref{q_max} for three different particle sizes, corresponding
to a size ratio of $\Phi=0.5,0.67$ and $0.83$ and a concentration of small
particles of 50\%. For smaller particles, the final amount of segregation is
higher for nearly all filling ratios of the cylinder. This can be explained by
the higher mobility of the small particles which traverse through the network
of voids of the large particles. 
%
\ifx\grdraft\undefined  \else
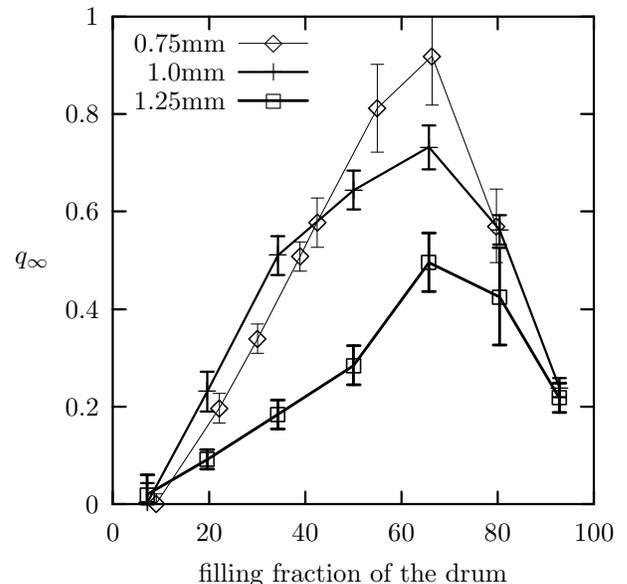
\begin{figure}[t]
\noindent	
{\centering 
\begin{picture}(1500,900)(0,100)
\tenrm
\thicklines \path(199,179)(219,179)
\thicklines \path(954,179)(934,179)
\put(177,179){\makebox(0,0)[r]{0}}
\thicklines \path(199,332)(219,332)
\thicklines \path(954,332)(934,332)
\put(177,332){\makebox(0,0)[r]{0.2}}
\thicklines \path(199,485)(219,485)
\thicklines \path(954,485)(934,485)
\put(177,485){\makebox(0,0)[r]{0.4}}
\thicklines \path(199,639)(219,639)
\thicklines \path(954,639)(934,639)
\put(177,639){\makebox(0,0)[r]{0.6}}
\thicklines \path(199,792)(219,792)
\thicklines \path(954,792)(934,792)
\put(177,792){\makebox(0,0)[r]{0.8}}
\thicklines \path(199,945)(219,945)
\thicklines \path(954,945)(934,945)
\put(177,945){\makebox(0,0)[r]{1}}
\thicklines \path(199,179)(199,199)
\thicklines \path(199,945)(199,925)
\put(199,134){\makebox(0,0){0}}
\thicklines \path(350,179)(350,199)
\thicklines \path(350,945)(350,925)
\put(350,134){\makebox(0,0){20}}
\thicklines \path(501,179)(501,199)
\thicklines \path(501,945)(501,925)
\put(501,134){\makebox(0,0){40}}
\thicklines \path(652,179)(652,199)
\thicklines \path(652,945)(652,925)
\put(652,134){\makebox(0,0){60}}
\thicklines \path(803,179)(803,199)
\thicklines \path(803,945)(803,925)
\put(803,134){\makebox(0,0){80}}
\thicklines \path(954,179)(954,199)
\thicklines \path(954,945)(954,925)
\put(954,134){\makebox(0,0){100}}
\thicklines \path(199,179)(954,179)(954,945)(199,945)(199,179)
\put(45,562){\makebox(0,0)[l]{\shortstack{$q_\infty$}}}
\put(576,67){\makebox(0,0){filling fraction of the drum}}
\put(375,903){\makebox(0,0)[r]{0.75mm}}
\thinlines \path(397,903)(505,903)
\thinlines \path(267,180)(267,180)(366,330)(426,439)(493,568)(520,622)(614,801)(700,883)(801,616)
\put(267,180){\raisebox{-1.2pt}{\makebox(0,0){$\Diamond$}}}
\put(366,330){\raisebox{-1.2pt}{\makebox(0,0){$\Diamond$}}}
\put(426,439){\raisebox{-1.2pt}{\makebox(0,0){$\Diamond$}}}
\put(493,568){\raisebox{-1.2pt}{\makebox(0,0){$\Diamond$}}}
\put(520,622){\raisebox{-1.2pt}{\makebox(0,0){$\Diamond$}}}
\put(614,801){\raisebox{-1.2pt}{\makebox(0,0){$\Diamond$}}}
\put(700,883){\raisebox{-1.2pt}{\makebox(0,0){$\Diamond$}}}
\put(801,616){\raisebox{-1.2pt}{\makebox(0,0){$\Diamond$}}}
\put(451,903){\raisebox{-1.2pt}{\makebox(0,0){$\Diamond$}}}
\thinlines \path(267,179)(267,195)
\thinlines \path(257,179)(277,179)
\thinlines \path(257,195)(277,195)
\thinlines \path(366,307)(366,353)
\thinlines \path(356,307)(376,307)
\thinlines \path(356,353)(376,353)
\thinlines \path(426,416)(426,462)
\thinlines \path(416,416)(436,416)
\thinlines \path(416,462)(436,462)
\thinlines \path(493,545)(493,591)
\thinlines \path(483,545)(503,545)
\thinlines \path(483,591)(503,591)
\thinlines \path(520,583)(520,660)
\thinlines \path(510,583)(530,583)
\thinlines \path(510,660)(530,660)
\thinlines \path(614,732)(614,870)
\thinlines \path(604,732)(624,732)
\thinlines \path(604,870)(624,870)
\thinlines \path(700,806)(700,945)
\thinlines \path(690,806)(710,806)
\thinlines \path(690,945)(710,945)
\thinlines \path(801,559)(801,674)
\thinlines \path(791,559)(811,559)
\thinlines \path(791,674)(811,674)
\put(267,180){\raisebox{-1.2pt}{\makebox(0,0){$\Diamond$}}}
\put(366,330){\raisebox{-1.2pt}{\makebox(0,0){$\Diamond$}}}
\put(426,439){\raisebox{-1.2pt}{\makebox(0,0){$\Diamond$}}}
\put(493,568){\raisebox{-1.2pt}{\makebox(0,0){$\Diamond$}}}
\put(520,622){\raisebox{-1.2pt}{\makebox(0,0){$\Diamond$}}}
\put(614,801){\raisebox{-1.2pt}{\makebox(0,0){$\Diamond$}}}
\put(700,883){\raisebox{-1.2pt}{\makebox(0,0){$\Diamond$}}}
\put(801,616){\raisebox{-1.2pt}{\makebox(0,0){$\Diamond$}}}
\put(375,858){\makebox(0,0)[r]{1.0mm}}
\thicklines \path(397,858)(505,858)
\thicklines \path(253,181)(253,181)(347,356)(458,570)(577,672)(695,740)(806,610)(900,362)
\put(253,181){\makebox(0,0){$+$}}
\put(347,356){\makebox(0,0){$+$}}
\put(458,570){\makebox(0,0){$+$}}
\put(577,672){\makebox(0,0){$+$}}
\put(695,740){\makebox(0,0){$+$}}
\put(806,610){\makebox(0,0){$+$}}
\put(900,362){\makebox(0,0){$+$}}
\put(451,858){\makebox(0,0){$+$}}
\thicklines \path(253,179)(253,212)
\thicklines \path(243,179)(263,179)
\thicklines \path(243,212)(263,212)
\thicklines \path(347,325)(347,387)
\thicklines \path(337,325)(357,325)
\thicklines \path(337,387)(357,387)
\thicklines \path(458,539)(458,600)
\thicklines \path(448,539)(468,539)
\thicklines \path(448,600)(468,600)
\thicklines \path(577,642)(577,703)
\thicklines \path(567,642)(587,642)
\thicklines \path(567,703)(587,703)
\thicklines \path(695,705)(695,774)
\thicklines \path(685,705)(705,705)
\thicklines \path(685,774)(705,774)
\thicklines \path(806,587)(806,633)
\thicklines \path(796,587)(816,587)
\thicklines \path(796,633)(816,633)
\thicklines \path(900,347)(900,377)
\thicklines \path(890,347)(910,347)
\thicklines \path(890,377)(910,377)
\put(253,181){\makebox(0,0){$+$}}
\put(347,356){\makebox(0,0){$+$}}
\put(458,570){\makebox(0,0){$+$}}
\put(577,672){\makebox(0,0){$+$}}
\put(695,740){\makebox(0,0){$+$}}
\put(806,610){\makebox(0,0){$+$}}
\put(900,362){\makebox(0,0){$+$}}
\put(375,813){\makebox(0,0)[r]{1.25mm}}
\Thicklines \path(397,813)(505,813)
\Thicklines \path(253,194)(253,194)(347,249)(458,320)(577,397)(695,559)(806,505)(900,346)
\put(253,194){\raisebox{-1.2pt}{\makebox(0,0){$\Box$}}}
\put(347,249){\raisebox{-1.2pt}{\makebox(0,0){$\Box$}}}
\put(458,320){\raisebox{-1.2pt}{\makebox(0,0){$\Box$}}}
\put(577,397){\raisebox{-1.2pt}{\makebox(0,0){$\Box$}}}
\put(695,559){\raisebox{-1.2pt}{\makebox(0,0){$\Box$}}}
\put(806,505){\raisebox{-1.2pt}{\makebox(0,0){$\Box$}}}
\put(900,346){\raisebox{-1.2pt}{\makebox(0,0){$\Box$}}}
\put(451,813){\raisebox{-1.2pt}{\makebox(0,0){$\Box$}}}
\Thicklines \path(253,179)(253,225)
\Thicklines \path(243,179)(263,179)
\Thicklines \path(243,225)(263,225)
\Thicklines \path(347,234)(347,265)
\Thicklines \path(337,234)(357,234)
\Thicklines \path(337,265)(357,265)
\Thicklines \path(458,297)(458,343)
\Thicklines \path(448,297)(468,297)
\Thicklines \path(448,343)(468,343)
\Thicklines \path(577,367)(577,428)
\Thicklines \path(567,367)(587,367)
\Thicklines \path(567,428)(587,428)
\Thicklines \path(695,513)(695,605)
\Thicklines \path(685,513)(705,513)
\Thicklines \path(685,605)(705,605)
\Thicklines \path(806,429)(806,582)
\Thicklines \path(796,429)(816,429)
\Thicklines \path(796,582)(816,582)
\Thicklines \path(900,323)(900,369)
\Thicklines \path(890,323)(910,323)
\Thicklines \path(890,369)(910,369)
\put(253,194){\raisebox{-1.2pt}{\makebox(0,0){$\Box$}}}
\put(347,249){\raisebox{-1.2pt}{\makebox(0,0){$\Box$}}}
\put(458,320){\raisebox{-1.2pt}{\makebox(0,0){$\Box$}}}
\put(577,397){\raisebox{-1.2pt}{\makebox(0,0){$\Box$}}}
\put(695,559){\raisebox{-1.2pt}{\makebox(0,0){$\Box$}}}
\put(806,505){\raisebox{-1.2pt}{\makebox(0,0){$\Box$}}}
\put(900,346){\raisebox{-1.2pt}{\makebox(0,0){$\Box$}}}
\end{picture} }
 \caption{Final amount of segregation for a concentration of 50\%
of small particles and three different size ratios.}
\label{q_max}
\end{figure}
\fi
Since the geometrical mixing will cause also mixing of the segregated core of
small particles with the large particles, best segregation should be achieved
for zero mixing, i.e. a filling fraction of 50\% when mono-disperse particles
are used, see also Sec.~\ref{4a}. In our case where we have a fluidized layer
with finite width,  the best segregation occurs not for a half-filled
cylinder, instead it occurs for a cylinder where the solid block under the
fluidized layer is 50\%.  The fluidized layer has a width of about three to
four particles and hence the total filling fraction for the least geometrical
mixing would be for a filling fraction of 60\%, which is in agreement with our
simulations (see Fig.~\ref{q_max}).

Concerning different concentrations of small particles, our parameter was
chosen in such a way that the final amount of segregation should be independent
of the volume fraction of small particles in the cylinder. We checked this
numerically for a concentration of 50\% and 33\% and  found a perfect agreement
within the error bars.
%
\ifx\grdraft\undefined  \else
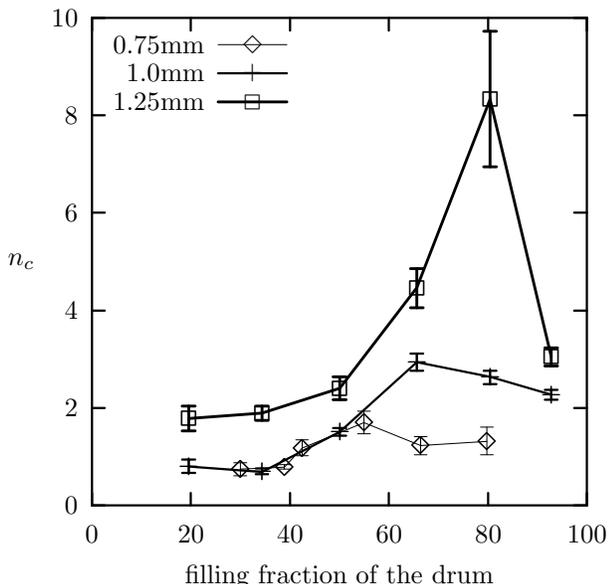
\begin{figure}
\noindent {\centering 
\begin{picture}(1500,900)(0,100)
\tenrm
\thicklines \path(177,179)(197,179)
\thicklines \path(954,179)(934,179)
\put(155,179){\makebox(0,0)[r]{0}}
\thicklines \path(177,332)(197,332)
\thicklines \path(954,332)(934,332)
\put(155,332){\makebox(0,0)[r]{2}}
\thicklines \path(177,485)(197,485)
\thicklines \path(954,485)(934,485)
\put(155,485){\makebox(0,0)[r]{4}}
\thicklines \path(177,639)(197,639)
\thicklines \path(954,639)(934,639)
\put(155,639){\makebox(0,0)[r]{6}}
\thicklines \path(177,792)(197,792)
\thicklines \path(954,792)(934,792)
\put(155,792){\makebox(0,0)[r]{8}}
\thicklines \path(177,945)(197,945)
\thicklines \path(954,945)(934,945)
\put(155,945){\makebox(0,0)[r]{10}}
\thicklines \path(177,179)(177,199)
\thicklines \path(177,945)(177,925)
\put(177,134){\makebox(0,0){0}}
\thicklines \path(332,179)(332,199)
\thicklines \path(332,945)(332,925)
\put(332,134){\makebox(0,0){20}}
\thicklines \path(488,179)(488,199)
\thicklines \path(488,945)(488,925)
\put(488,134){\makebox(0,0){40}}
\thicklines \path(643,179)(643,199)
\thicklines \path(643,945)(643,925)
\put(643,134){\makebox(0,0){60}}
\thicklines \path(799,179)(799,199)
\thicklines \path(799,945)(799,925)
\put(799,134){\makebox(0,0){80}}
\thicklines \path(954,179)(954,199)
\thicklines \path(954,945)(954,925)
\put(954,134){\makebox(0,0){100}}
\thicklines \path(177,179)(954,179)(954,945)(177,945)(177,179)
\put(45,562){\makebox(0,0)[l]{\shortstack{$n_c$}}}
\put(565,67){\makebox(0,0){filling fraction of the drum}}
\put(353,903){\makebox(0,0)[r]{0.75mm}}
\thinlines \path(375,903)(483,903)
\thinlines \path(410,236)(410,236)(479,239)(507,269)(604,309)(693,273)(797,280)
\put(410,236){\raisebox{-1.2pt}{\makebox(0,0){$\Diamond$}}}
\put(479,239){\raisebox{-1.2pt}{\makebox(0,0){$\Diamond$}}}
\put(507,269){\raisebox{-1.2pt}{\makebox(0,0){$\Diamond$}}}
\put(604,309){\raisebox{-1.2pt}{\makebox(0,0){$\Diamond$}}}
\put(693,273){\raisebox{-1.2pt}{\makebox(0,0){$\Diamond$}}}
\put(797,280){\raisebox{-1.2pt}{\makebox(0,0){$\Diamond$}}}
\put(429,903){\raisebox{-1.2pt}{\makebox(0,0){$\Diamond$}}}
\thinlines \path(410,226)(410,246)
\thinlines \path(400,226)(420,226)
\thinlines \path(400,246)(420,246)
\thinlines \path(479,236)(479,243)
\thinlines \path(469,236)(489,236)
\thinlines \path(469,243)(489,243)
\thinlines \path(507,257)(507,282)
\thinlines \path(497,257)(517,257)
\thinlines \path(497,282)(517,282)
\thinlines \path(604,292)(604,327)
\thinlines \path(594,292)(614,292)
\thinlines \path(594,327)(614,327)
\thinlines \path(693,259)(693,287)
\thinlines \path(683,259)(703,259)
\thinlines \path(683,287)(703,287)
\thinlines \path(797,259)(797,302)
\thinlines \path(787,259)(807,259)
\thinlines \path(787,302)(807,302)
\put(410,236){\raisebox{-1.2pt}{\makebox(0,0){$\Diamond$}}}
\put(479,239){\raisebox{-1.2pt}{\makebox(0,0){$\Diamond$}}}
\put(507,269){\raisebox{-1.2pt}{\makebox(0,0){$\Diamond$}}}
\put(604,309){\raisebox{-1.2pt}{\makebox(0,0){$\Diamond$}}}
\put(693,273){\raisebox{-1.2pt}{\makebox(0,0){$\Diamond$}}}
\put(797,280){\raisebox{-1.2pt}{\makebox(0,0){$\Diamond$}}}
\put(353,858){\makebox(0,0)[r]{1.0mm}}
\thicklines \path(375,858)(483,858)
\thicklines \path(329,240)(329,240)(444,232)(566,295)(687,404)(802,381)(898,353)
\put(329,240){\makebox(0,0){$+$}}
\put(444,232){\makebox(0,0){$+$}}
\put(566,295){\makebox(0,0){$+$}}
\put(687,404){\makebox(0,0){$+$}}
\put(802,381){\makebox(0,0){$+$}}
\put(898,353){\makebox(0,0){$+$}}
\put(429,858){\makebox(0,0){$+$}}
\thicklines \path(329,230)(329,251)
\thicklines \path(319,230)(339,230)
\thicklines \path(319,251)(339,251)
\thicklines \path(444,228)(444,237)
\thicklines \path(434,228)(454,228)
\thicklines \path(434,237)(454,237)
\thicklines \path(566,289)(566,301)
\thicklines \path(556,289)(576,289)
\thicklines \path(556,301)(576,301)
\thicklines \path(687,391)(687,418)
\thicklines \path(677,391)(697,391)
\thicklines \path(677,418)(697,418)
\thicklines \path(802,370)(802,391)
\thicklines \path(792,370)(812,370)
\thicklines \path(792,391)(812,391)
\thicklines \path(898,345)(898,361)
\thicklines \path(888,345)(908,345)
\thicklines \path(888,361)(908,361)
\put(329,240){\makebox(0,0){$+$}}
\put(444,232){\makebox(0,0){$+$}}
\put(566,295){\makebox(0,0){$+$}}
\put(687,404){\makebox(0,0){$+$}}
\put(802,381){\makebox(0,0){$+$}}
\put(898,353){\makebox(0,0){$+$}}
\put(353,813){\makebox(0,0)[r]{1.25mm}}
\Thicklines \path(375,813)(483,813)
\Thicklines \path(329,316)(329,316)(444,324)(566,363)(687,521)(802,817)(898,413)
\put(329,316){\raisebox{-1.2pt}{\makebox(0,0){$\Box$}}}
\put(444,324){\raisebox{-1.2pt}{\makebox(0,0){$\Box$}}}
\put(566,363){\raisebox{-1.2pt}{\makebox(0,0){$\Box$}}}
\put(687,521){\raisebox{-1.2pt}{\makebox(0,0){$\Box$}}}
\put(802,817){\raisebox{-1.2pt}{\makebox(0,0){$\Box$}}}
\put(898,413){\raisebox{-1.2pt}{\makebox(0,0){$\Box$}}}
\put(429,813){\raisebox{-1.2pt}{\makebox(0,0){$\Box$}}}
\Thicklines \path(329,296)(329,335)
\Thicklines \path(319,296)(339,296)
\Thicklines \path(319,335)(339,335)
\Thicklines \path(444,313)(444,335)
\Thicklines \path(434,313)(454,313)
\Thicklines \path(434,335)(454,335)
\Thicklines \path(566,345)(566,381)
\Thicklines \path(556,345)(576,345)
\Thicklines \path(556,381)(576,381)
\Thicklines \path(687,490)(687,551)
\Thicklines \path(677,490)(697,490)
\Thicklines \path(677,551)(697,551)
\Thicklines \path(802,711)(802,924)
\Thicklines \path(792,711)(812,711)
\Thicklines \path(792,924)(812,924)
\Thicklines \path(898,398)(898,427)
\Thicklines \path(888,398)(908,398)
\Thicklines \path(888,427)(908,427)
\put(329,316){\raisebox{-1.2pt}{\makebox(0,0){$\Box$}}}
\put(444,324){\raisebox{-1.2pt}{\makebox(0,0){$\Box$}}}
\put(566,363){\raisebox{-1.2pt}{\makebox(0,0){$\Box$}}}
\put(687,521){\raisebox{-1.2pt}{\makebox(0,0){$\Box$}}}
\put(802,817){\raisebox{-1.2pt}{\makebox(0,0){$\Box$}}}
\put(898,413){\raisebox{-1.2pt}{\makebox(0,0){$\Box$}}}
\end{picture} }
\caption{Characteristic number of revolutions for segregation 
for a concentration of 50\%
of small particles.}
\label{q_time}
\end{figure}
\fi
The speed of the segregation is characterized by $t_c$, see Eq.~(\ref{fit}), 
stating that for $t=t_c$ the system has reached a segregation of 63\% of the
final value of $q_\infty$.  In Fig.~\ref{q_time}, we plot the characteristic
number of revolutions, defined as $n_c=\frac{\Omega t_c}{2\pi}$,  as a function
of the volume filling fraction of the cylinder for three different sizes of the
small  particles corresponding to particle size ratios of $\Phi=0.50,0.67$ and
$0.83$ again. The segregation times are sufficiently smaller for the smaller
particles  which is also found in experiments~\cite{williams76}. They also show
a general trend of increasing with increasing filling fraction.   However, for
a more than half-filled drum,   where the exact value depends on the particle
size due to the different width of the fluidized layers, the segregation
becomes faster again. This can be explained as follows: Due to the nature of
the geometrical mixing, an  unmixed core will persists in the middle of the
cylinder for a filling fraction greater than 50\%. The exact number depends on
the width of the fluidized layer which depends e.g. on the rotation speed of
the cylinder. Close to a filling fraction  of one, only a small ring close to
the cylinder wall can participate in the segregation process. Consequently, the
final amount of segregation is small which agrees with Fig.~\ref{q_max}, but
this value is reached fast, therefore the segregation time, $t_c$, is small.
The numerical data  indicates that the filling fraction which corresponds to
the maximal value of $t_c$ decreases with decreasing particle size.

It is well known, that the segregation process is faster and more pronounced if
the particle size ratio becomes smaller~\cite{williams76}. The results from a
two dimensional rotating drum model indicate that segregation  is observed for
an arbitrary small size ratio~\cite{baumann94}, whereas the data from vertical
shaking experiments suggest a cut-off ratio around $\Phi=0.5$~\cite{vanel97}.
In order to address this question, we show in Fig.~\ref{critical} the final
amount of segregation, $q_\infty$, as a function of the particle size ratio,
$\Phi$, for a concentration of small particles and a volume filling fraction of
50\%. Even though obtaining accurate data for values  of $\Phi$ close to one is
rather difficult due to the long segregation time, the data shown in
Fig.~\ref{critical} support the hypothesis from~\cite{baumann94} that
segregation will be present for any finite size difference. This was determined
by using a linear fit of the form
\begin{equation}
q_\infty(\Phi) = c ( 1 - \Phi_0)
\end{equation}
which gives $c =1.6\pm 0.1$  and $\Phi_0 = 1.00 \pm 0.02$ when all seven data
points are used for the fit;  shown as a dashed line in Fig.~\ref{critical}.
%
\ifx\grdraft\undefined  \else
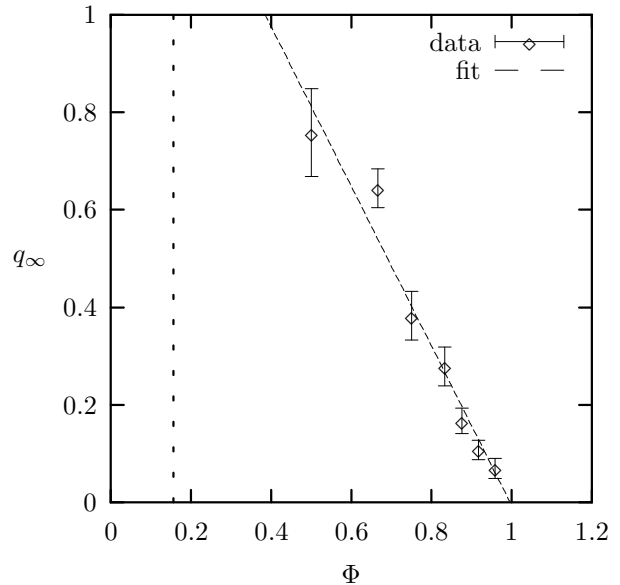
\begin{figure}
\noindent	
{\centering 
\begin{picture}(1500,900)(0,50)
\tenrm
\thicklines \path(199,179)(219,179)
\thicklines \path(954,179)(934,179)
\put(177,179){\makebox(0,0)[r]{0}}
\thicklines \path(199,332)(219,332)
\thicklines \path(954,332)(934,332)
\put(177,332){\makebox(0,0)[r]{0.2}}
\thicklines \path(199,485)(219,485)
\thicklines \path(954,485)(934,485)
\put(177,485){\makebox(0,0)[r]{0.4}}
\thicklines \path(199,639)(219,639)
\thicklines \path(954,639)(934,639)
\put(177,639){\makebox(0,0)[r]{0.6}}
\thicklines \path(199,792)(219,792)
\thicklines \path(954,792)(934,792)
\put(177,792){\makebox(0,0)[r]{0.8}}
\thicklines \path(199,945)(219,945)
\thicklines \path(954,945)(934,945)
\put(177,945){\makebox(0,0)[r]{1}}
\thicklines \path(199,179)(199,199)
\thicklines \path(199,945)(199,925)
\put(199,134){\makebox(0,0){0}}
\thicklines \path(325,179)(325,199)
\thicklines \path(325,945)(325,925)
\put(325,134){\makebox(0,0){0.2}}
\dashline{10}(297,179)(297,945)
\thicklines \path(451,179)(451,199)
\thicklines \path(451,945)(451,925)
\put(451,134){\makebox(0,0){0.4}}
\thicklines \path(577,179)(577,199)
\thicklines \path(577,945)(577,925)
\put(577,134){\makebox(0,0){0.6}}
\thicklines \path(702,179)(702,199)
\thicklines \path(702,945)(702,925)
\put(702,134){\makebox(0,0){0.8}}
\thicklines \path(828,179)(828,199)
\thicklines \path(828,945)(828,925)
\put(828,134){\makebox(0,0){1}}
\thicklines \path(954,179)(954,199)
\thicklines \path(954,945)(954,925)
\put(954,134){\makebox(0,0){1.2}}
\thicklines \path(199,179)(954,179)(954,945)(199,945)(199,179)
\put(45,562){\makebox(0,0)[l]{\shortstack{$q_\infty$}}}
\put(576,67){\makebox(0,0){$\Phi$}}
\put(780,903){\makebox(0,0)[r]{data}}
\thinlines \path(802,903)(910,903)
\thinlines \path(802,913)(802,893)
\thinlines \path(910,913)(910,893)
\thinlines \path(514,691)(514,829)
\thinlines \path(504,691)(524,691)
\thinlines \path(504,829)(524,829)
\thinlines \path(618,642)(618,703)
\thinlines \path(608,642)(628,642)
\thinlines \path(608,703)(628,703)
\thinlines \path(671,434)(671,511)
\thinlines \path(661,434)(681,434)
\thinlines \path(661,511)(681,511)
\thinlines \path(723,362)(723,423)
\thinlines \path(713,362)(733,362)
\thinlines \path(713,423)(733,423)
\thinlines \path(750,287)(750,327)
\thinlines \path(740,287)(760,287)
\thinlines \path(740,327)(760,327)
\thinlines \path(776,246)(776,277)
\thinlines \path(766,246)(786,246)
\thinlines \path(766,277)(786,277)
\thinlines \path(802,217)(802,248)
\thinlines \path(792,217)(812,217)
\thinlines \path(792,248)(812,248)
\put(514,760){\raisebox{-1.2pt}{\makebox(0,0){$\diamond$}}}
\put(618,672){\raisebox{-1.2pt}{\makebox(0,0){$\diamond$}}}
\put(671,472){\raisebox{-1.2pt}{\makebox(0,0){$\diamond$}}}
\put(723,393){\raisebox{-1.2pt}{\makebox(0,0){$\diamond$}}}
\put(750,307){\raisebox{-1.2pt}{\makebox(0,0){$\diamond$}}}
\put(776,262){\raisebox{-1.2pt}{\makebox(0,0){$\diamond$}}}
\put(802,233){\raisebox{-1.2pt}{\makebox(0,0){$\diamond$}}}
\put(856,903){\raisebox{-1.2pt}{\makebox(0,0){$\diamond$}}}
\put(780,858){\makebox(0,0)[r]{fit}}
\thinlines \drawline[-50](802,858)(910,858)
\thinlines \drawline[-50](441,945)(443,941)(451,926)(458,911)(466,896)(474,881)(481,865)(489,850)(496,835)(504,820)(512,805)(519,789)(527,774)(535,759)(542,744)(550,729)(557,713)(565,698)(573,683)(580,668)(588,653)(596,637)(603,622)(611,607)(618,592)(626,577)(634,561)(641,546)(649,531)(657,516)(664,501)(672,485)(679,470)(687,455)(695,440)(702,425)(710,410)(718,394)(725,379)(733,364)(740,349)(748,334)(756,318)(763,303)(771,288)(779,273)(786,258)(794,242)(801,227)(809,212)(817,197)
\thinlines \drawline[-50](817,197)(824,182)(826,179)
\end{picture} }
 \caption{Final amount of segregation as function of the particle size ratio 
$\Phi=\frac{r}{R}$  (the vertical dashed line denotes $\Phi_T$).}
\label{critical}
\end{figure}
\fi
Due to our definition, the maximal achievable value for the final amount of
segregation is $q_\infty=1$. Therefore for small size ratios, the behavior must
deviate from the linear dependence which is already  visible for the value for
$\Phi=0.5$ which was obtained by interpolating between two filling fractions.
Obtaining data points for even lower values of $\Phi$ is nearly infeasible by
todays computers, due to the large demand on computer time caused by the large
particle numbers. For values  of $\Phi\leq\frac{2-\sqrt{3}}{\sqrt{3}}$ (wide
dashed line in Fig.~\ref{critical}) we  expect a completely different behavior
since the small particles are then sufficiently small to propagate through the
voids of a three--dimensional hexagonal packing. Please note that since we have
a random packing, see Fig.~\ref{density}, the threshold value should even be
higher and $\Phi_{T}:=\frac{2-\sqrt{3}}{\sqrt{3}}$ just serves as a lower
bound.

\subsection{Half--filled pre--set cylinder}
After having discussed the segregation dynamics as function of filling fraction
and  particle size ratio in the preceding paragraphs, we will now turn to
illustrate the interplay between  mixing and segregation. As already mentioned
in Sec.~\ref{4a}, no geometrical mixing was found for an exactly half-filled
drum in the discrete avalanche regime~\cite{metcalfe95,peratt96,dorogovtsev98}.
The continuous flow limit taken in Ref.~\cite{peratt96} suggests that this is
also true in the continuous flow regime but the numerical data sets given in 
Fig.~\ref{q_max} and~\ref{q_time} do not support this hypothesis which can be
attributed to the fact that the fluidized layer has a non-zero width.

In order to illustrate this point in more depth, we are starting with an
initial  configuration where the left half of the cylinder is purely composed
out of large beads  (R=1.5mm) and the  right half out of small beads (r=1.0mm) 
giving a total number of 4420 particles (see Fig.~\ref{split}, top left
picture).  After turning the drum counter-clockwise for 1.6 seconds at
$\Omega=15$rpm,  which would simple interchange the regions occupied by large
and small particles if no mixing would be present, the interface is still well
defined and nearly a straight  line (top right picture). After turning for 2.8
seconds,  the interface between the large and small beads is still quite
sharp,  albeit it is not a straight line anymore.   After the start of the
rotation, it takes 0.23 seconds for the continuous flow to set in and from
Sec.~\ref{4a}, we recall that it takes roughly 2.65 seconds for a particle to
undergo a full revolution. The tongue of small particles into the large ones at
the center of the cylinder is the starting point of forming a core of small
particles. After 
rotating for 5.3 seconds, which corresponds to two full particle revolutions,
the formation of a core of small (white) particles is even more pronounced. 
After two full particle revolutions, $t=7.8$s, the shape of the interface
between large and small particles  close to the cylinder wall  becomes even
more diffusive. It rather resembles a diffusion process along the azimuthal
direction which can be described in a similar fashion as the front propagation
along the axial direction in rotating cylinders~\cite{ristow98}. The
segregation mechanism in the pre--set cylinder starts immediately, i.e. {\em
no} mixing of the two components is necessary in order to obtain a radially
segregated core of small particles. The final picture in Fig.~\ref{split}
(bottom right) corresponds to 28 particle revolutions and shows a nearly
symmetric, well segregated cluster of small particles. Also note that hardly
any large particles are found in the segregated core of small particles whereas
smaller particles are still found close to the wall of the drum. We expect the
latter effect to disappear when the drum is rotated for long enough times.

In order to determine the degree of mixing in the horizontal cylinder along the
initially sharp vertical front we use a procedure proposed by Metcalfe et
al.~\cite{metcalfe95}. We calculate the center of mass for each particle size,
project it onto the free surface which is initially horizontal and calculate
the distance of each of the two centers of mass. 
\ifx\grdraft\undefined  \else
\begin{figure}[t]
\noindent	
\begin{minipage}[b]{.49\linewidth}	
\psfig{file=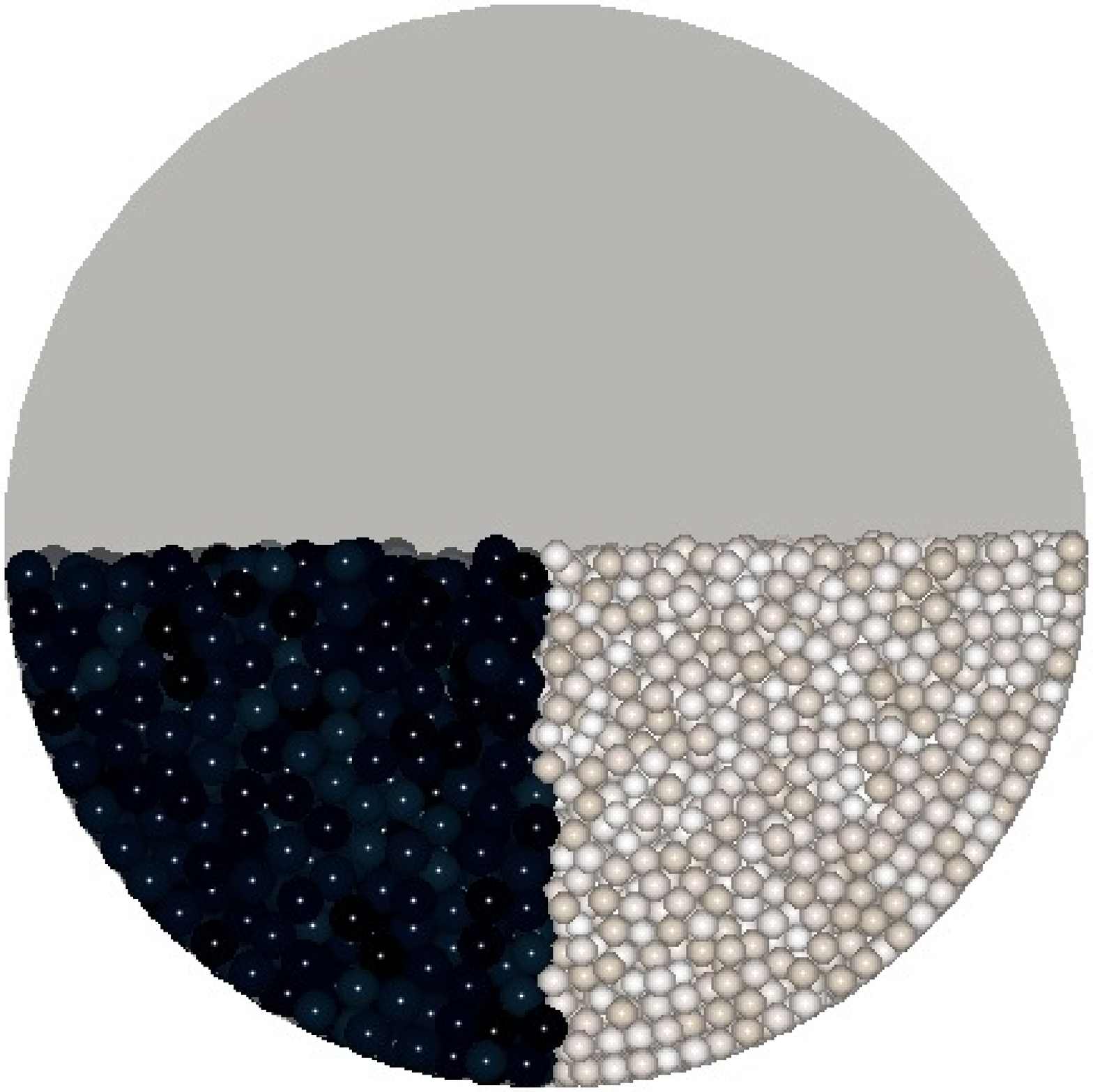,width=0.9\linewidth,angle=-90}\\
\centering \it 0 sec
\end{minipage} 
\begin{minipage}[b]{.49\linewidth}	
\psfig{file=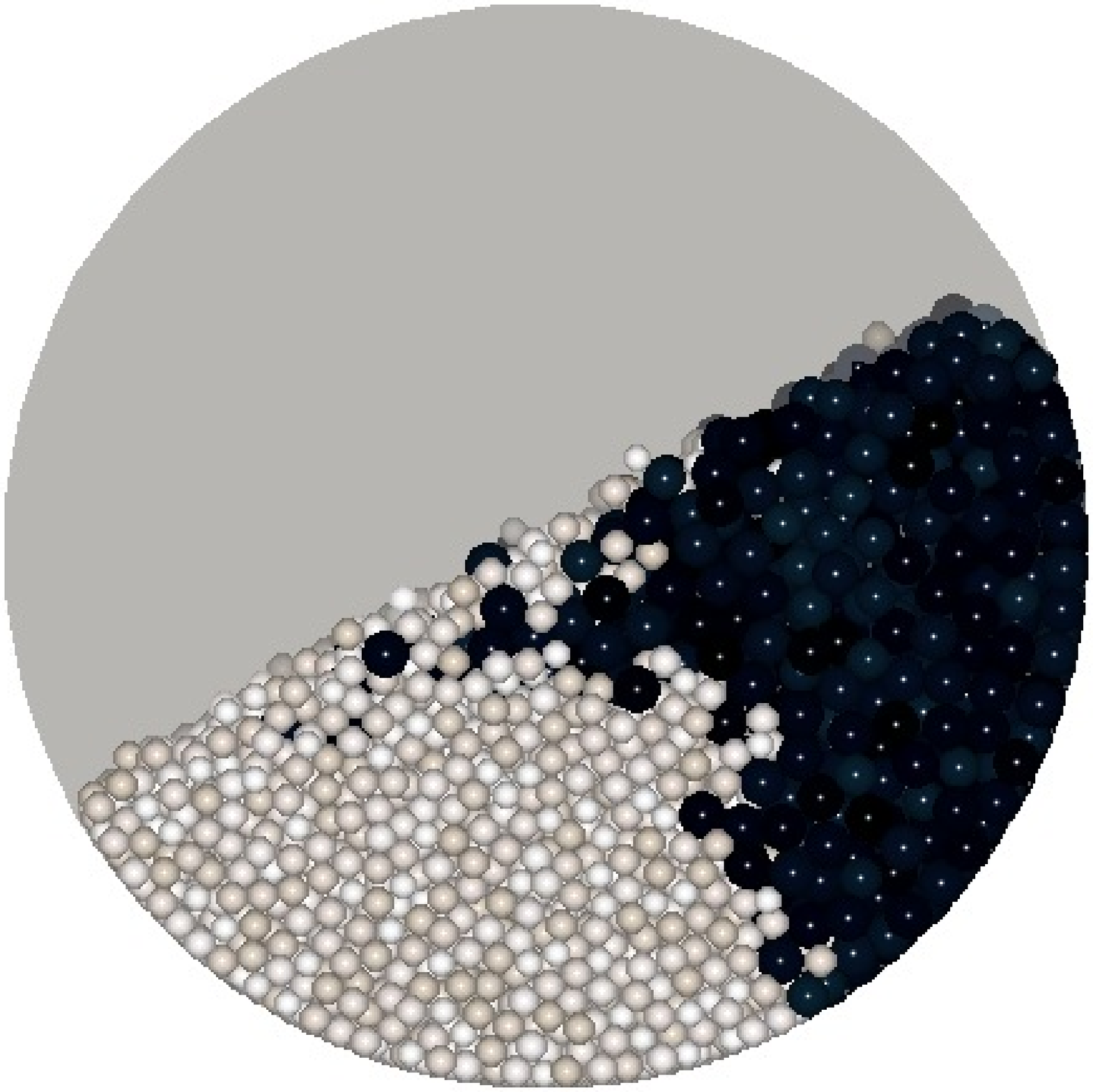,width=0.9\linewidth,angle=-90}\\
\centering \it 1.6 sec
\end{minipage} 
\begin{minipage}[b]{.49\linewidth}	
\psfig{file=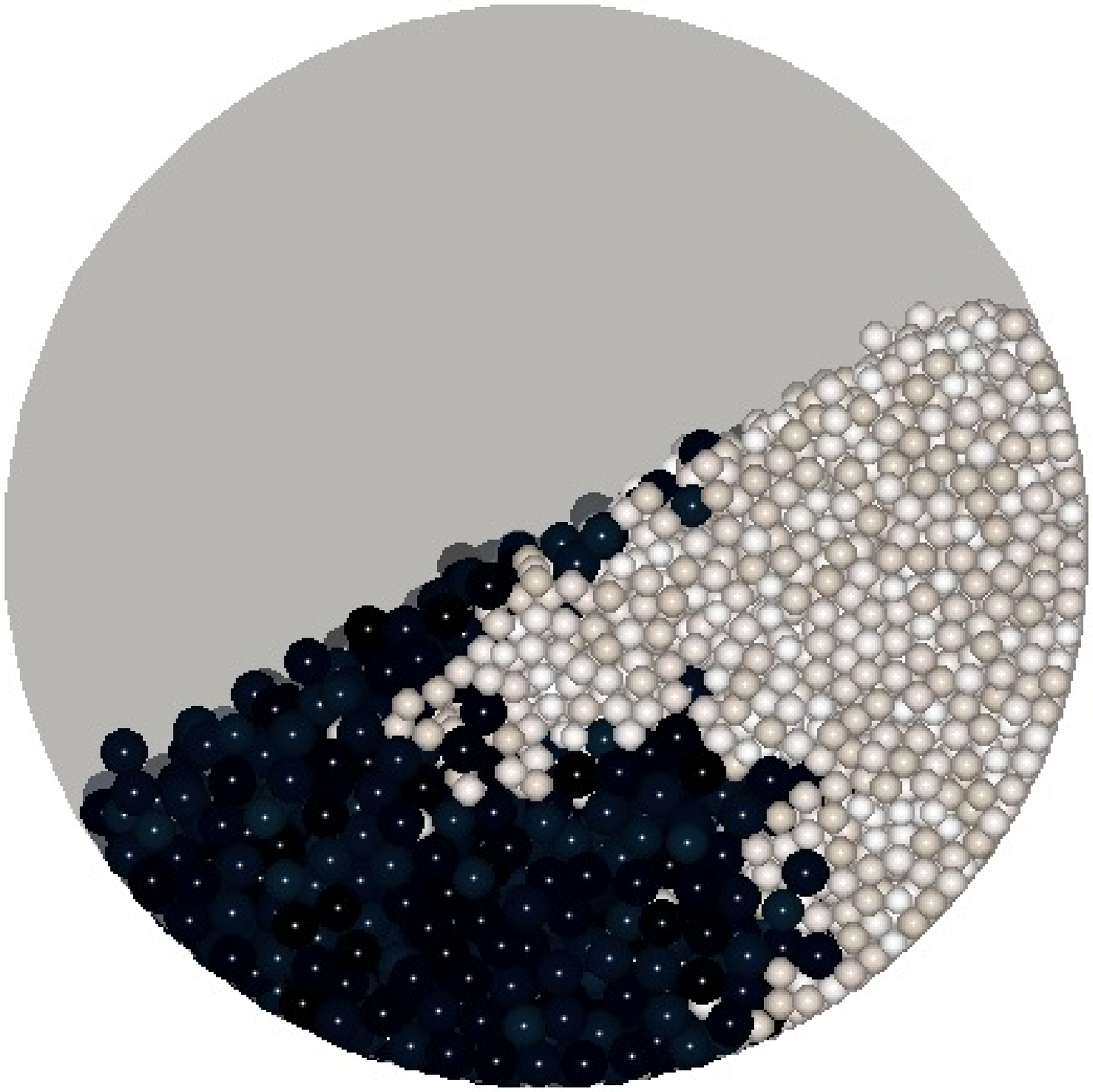,width=0.9\linewidth,angle=-90}\\
\centering \it 2.8 sec
\end{minipage} 
\begin{minipage}[b]{.49\linewidth}	
\psfig{file=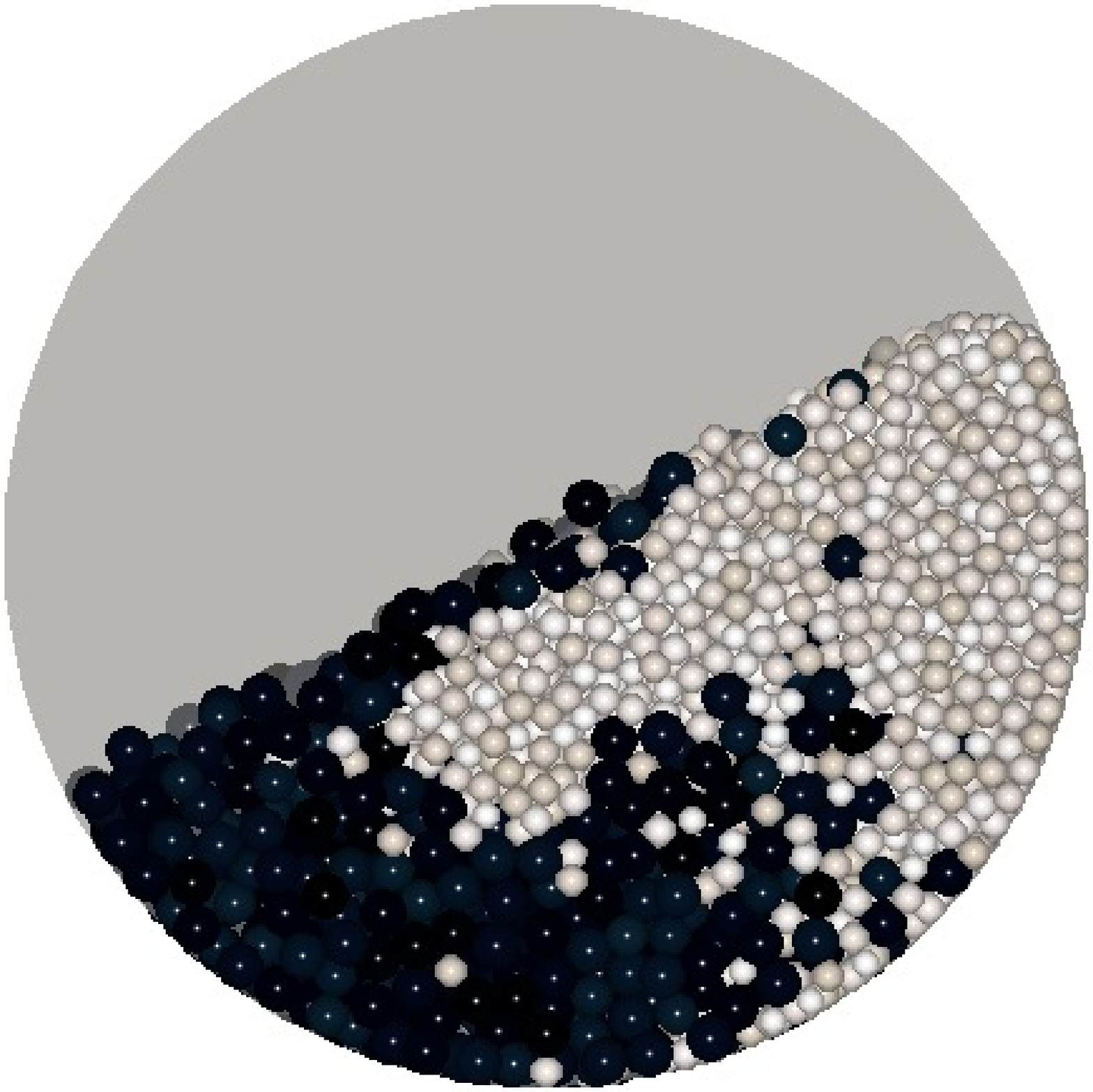,width=0.9\linewidth,angle=-90}\\
\centering \it 5.3 sec
\end{minipage} 
\begin{minipage}[b]{.49\linewidth}	
\psfig{file=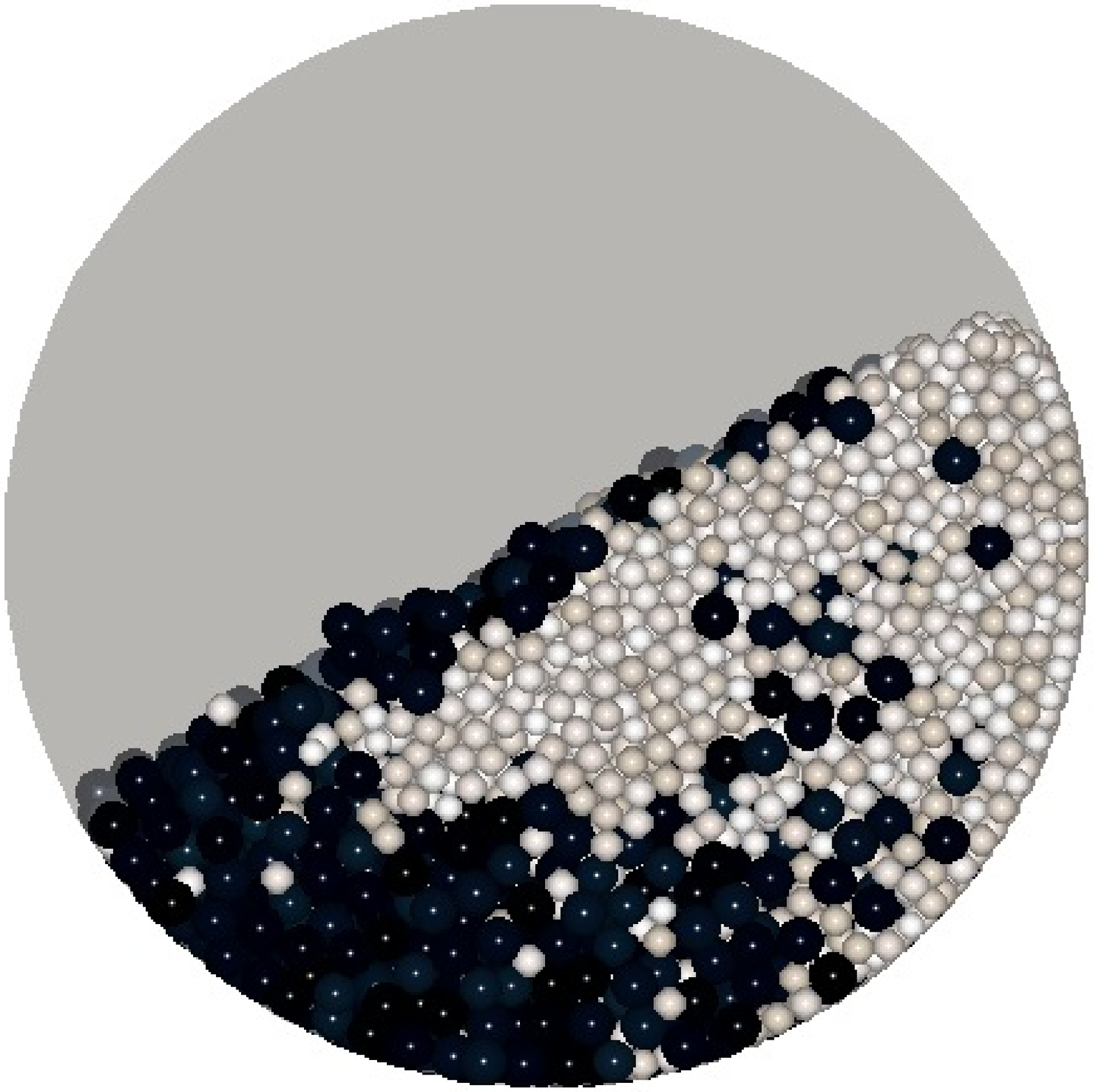,width=0.9\linewidth,angle=-90}\\
\centering \it 7.8 sec
\end{minipage} 
\begin{minipage}[b]{.49\linewidth}	
\psfig{file=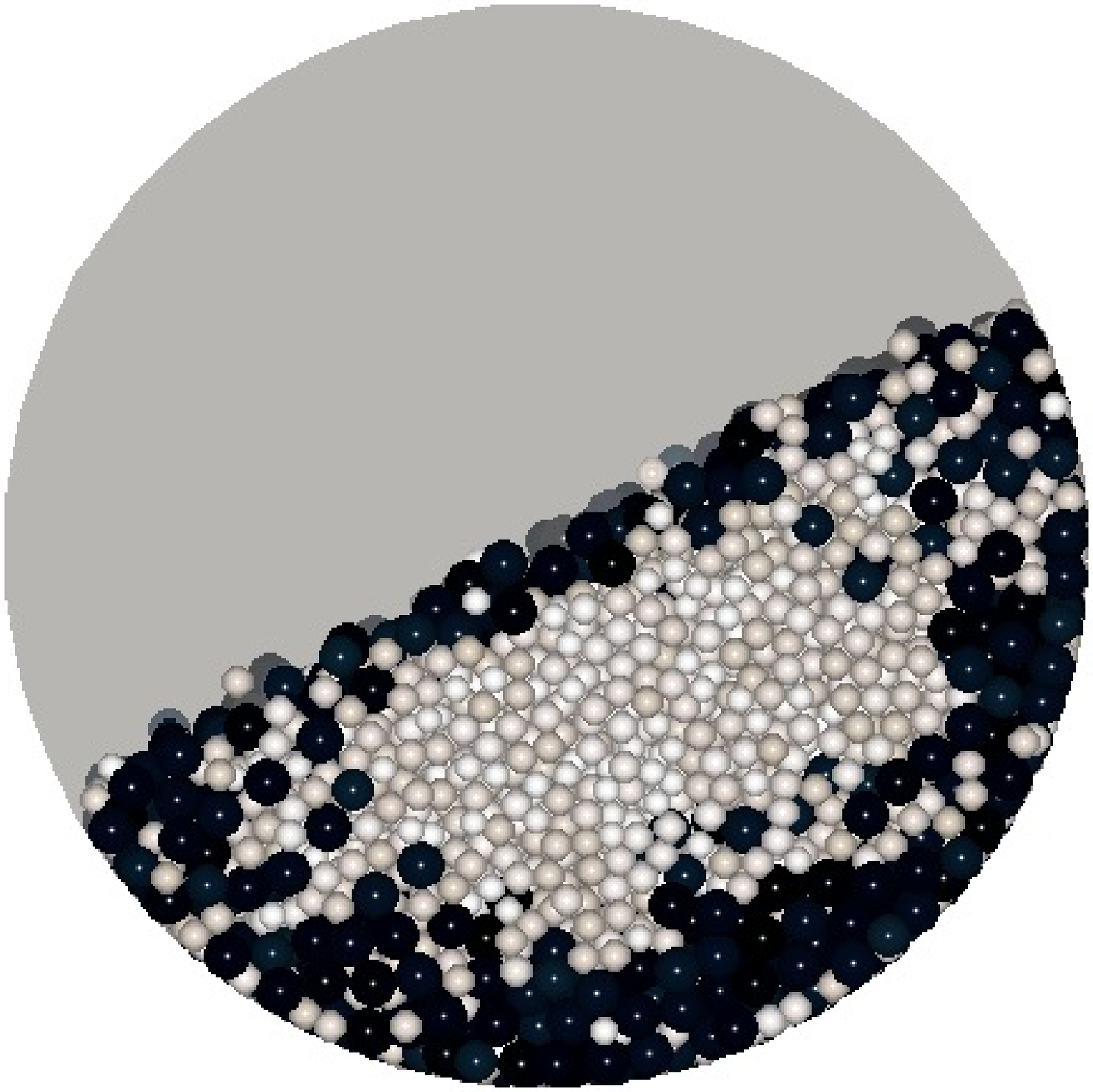,width=0.9\linewidth,angle=-90}\\
\centering \it 56.2 sec
\end{minipage} 
\caption{Different snapshots of the cylinder with a starting condition, where 
initially all the small (large)  particles are on the right (left) side of the
cylinder (particle radii 1.0mm and 1.5mm).}
\label{split}
\end{figure}
\fi
The time evolution of this distance, $\xi_c$, which was made dimensionless by
dividing by the distance of the start configuration is shown in
Fig.~\ref{centroid}. It corresponds to the configuration shown in
Fig.~\ref{split}, i.e. a half-filled cylinder containing an equal volume
fraction of $1$mm and $1.5$mm particles. In Ref.~\cite{metcalfe95}, this
procedure was used to show the mixing of mono-disperse particles in a rotating
drum at a filling fraction of $f=39\%$. Since geometrical mixing is observed
for this  filling fraction, the centroid positions decayed in time and could be
well approximated by
\begin{equation}
\xi_c(t) = \cos\left(\frac{2\pi t}{T}\right)e^{-t/t_c}
\label{decay}
\end{equation}
where $T$ stands for the period, see Eq.~(\ref{period}). On the contrary, no
geometrical mixing is observed for mono-disperse particles in a half-filled
drum which would correspond to  a characteristic time $t_c=\infty$ in
Eq.~(\ref{decay}) and lead to non-decaying oscillations. For a binary particle
mixture, the segregation process will lead to a decay of the distance of the
two centroids in time for {\em any} filling fraction and we have chosen to
present numerical results for  counter-clockwise rotation and $f=50\%$ in
Fig.~\ref{centroid} to illustrate this.
\ifx\grdraft\undefined  \else
\begin{figure}[t]
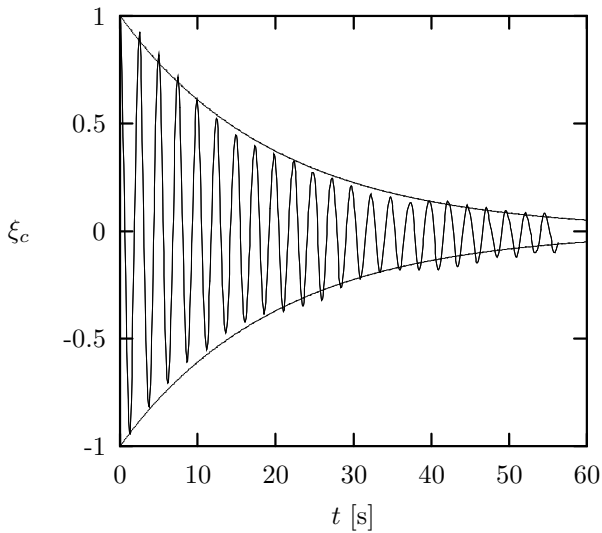

\noindent	
{\centering \input centroid.tex }
 \caption{Normalized centroid position projected onto the free surface
of a half-filled cylinder with 1.0mm and 1.5mm particles.}
\label{centroid}
\end{figure}
\fi

The numerical data can be well fitted by an exponentially decaying oscillation
according to Eq.~(\ref{decay}). This gives $T=2.48$s which is in excellent
agreement with  Fig.~\ref{q_fourier}, and values of $t_c=20.7$s for clockwise
rotation and $t_c=20.1$s for counter-clockwise rotation
which are the same within the error. 
The exponentially
decaying part was added to Fig.~\ref{centroid} as dashed line.  In the
beginning, $t<20$s a slight asymmetry is visible towards negative values which
was also observed in other numerical simulations using a two dimensional
geometrical  model~\cite{baumann97}. However, if this persists on the long run
and will led to a  final non-zero value for the centroid position could not be
determined by this procedure since the deviations from the exponential  fit
were usually higher than the calculated offset of $(0.004\pm 0.001)$cm. 
Since the numerical data seems to indicate that  a pure exponential decay is
too slow in the beginning and two fast for longer times we also tried a
stretched exponential decay of the form $e^{-(t/t_c)^\beta}$. This gives values
of $t_c=(14.0\pm1.8)s$ and $\beta=0.88\pm0.05$. However,  we can not rule out
that the deviation  of the exponential law may be due to the numerical noise
and a more general statistical theory is needed to resolve this question.

The procedure described above to follow the centroid dynamics is not capable to
determine the  depth of the centroid position below the free surface due to the
projection onto the  surface. However, the last picture of Fig.~\ref{split}
shows that the number of layers of  large particles below and above the
segregated cluster of small particles is not the same.  This leads to different
distances of the center of mass below the free surface for small and  large
particles. To determine its dynamic, we plot in Fig.~\ref{center} the Euclidean
distance of the two centroids. When properly shifted by $T/2$, the curves for
clockwise and counter-clockwise  rotations show a similar behavior where
fluctuations are a little less pronounced for the  latter case. The minima
correspond to configurations similar to the one shown in the bottom right
picture of Fig.~\ref{split} when the two centers of mass lie on a line which 
goes through the origin of the drum. Even though the minimal distance during
the time evolution is close to zero, a non-zero value of $0.22$cm is estimated
for the stationary state without oscillations.
\ifx\grdraft\undefined  \else
\begin{figure}[t]
\noindent	
\psfig{file=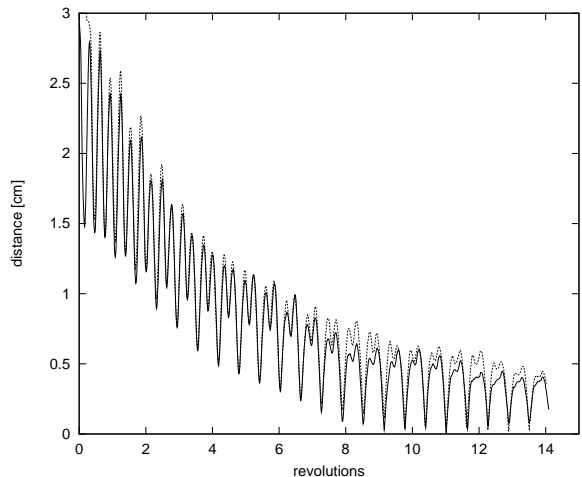,width=0.9\linewidth}\\
 \caption{Distance of the centers of a half--filled cylinder with 1.0mm and 
 1.5mm particles (\ -\ -\ - clockwise, --- counter-clockwise rotation).}
\label{center}
\end{figure}
\fi

\section{Conclusions}
The main conclusions of this paper are as follows. In the continuous flow
regime size segregation takes place for arbitrary small differences in particle
size. To quantify this result we introduced an appropriate order parameter,
which allowed us to compare directly all different drum scenarios. From this we
showed, that the radial segregation process is faster and more pronounced for
particles with a large size difference and that the final amount of segregation
shows a linear dependence  when approaching a particle size ratio of one; thus
no threshold value  for radial segregation exists, which was an unresolved
question for long time.

We also studied in detail the interplay between mixing and segregation in
rotating cylinders  for different volume filling fractions of the cylinder and
found that the highest achievable segregation can be obtained for a slightly
more than half-filled cylinder and therefore also least mixing. The difference
to a mono-disperse system in the discrete avalanche regime  could be attributed
to  the fluidized layer leading to a partial destruction of the underlying 
already segregated core where the destruction increases with layer width.  

When starting with an initial configuration that contains well--separated
regions of small and large particles  no mixing of the components is necessary
in order to obtain a radially segregated core. This inter-penetration process 
resembles a diffusion process and segregation starts immediately without
undergoing a previous mixing of the two particle components.


\section*{Acknowledgements}
We would like to thank the HLRZ in J\"ulich and the HRZ Marburg for supporting
us with a generous grant of computer-time on their Cray T3E and IBM SP2,
respectively. Financial support by the Deutsche Forschungsgemeinschaft is
greatfully acknowledged.

\section*{Appendix: Rolling Friction} 
One of the most distinct properties of a glass bead is its ability to roll,
therefore rolling had to be included into our model. The drawback was, that a
glass bead would have had a Coulomb friction of zero with our frictional laws;
i.e. a glass sphere would start to roll even on an infinitesimal inclined plane
or a particle would roll on forever on every flat plane, which in reality
clearly does not occur. For an ideal elastic particle, the deformation of the
particle would be symmetric to the point of contact and therefore the resulting
counter force of the plane would be exactly opposite to the gravitational force
$F_N$ for all times. In reality the deformation is not elastic, i.e. the
deformation lags behind as is indicated in Fig.~\ref{sphere} for a particle on
an ideal hard plane. The counter force of the plane $F$ acting on the particle
gets mediated by the deformation of the particle. The point where this force
acts on is shifted slightly by $r_0$ to the back, the normal component of $F$
compensates the gravitational force $F_N$ exactly (otherwise the particle would
bounce); leaving the tangential component of $F$ which acts as rolling
resistance which must be compensated by a dragging force for a particle with
constant velocity on a flat plane. Also simulation shows that the angle of
repose is much too small in comparison with experiment without rolling
resistance.  To overcome this weakness we add rolling
resistance~\cite{johnson85} to our model, see Fig.~\ref{sphere}, by using
\begin{equation}
  F_{r} = \frac{r_0}{R} F_{N} \quad .\label{rolling}
\end{equation}
Here the rolling resistance $r_0$ is a constant material  parameter and results
from the slight viscoelasticity of the  materials. $r_0$ is of the order of
$10^{-3}-10^{-5}$mm for most the materials. For particle-particle interactions,
we take the same law for rolling resistance as for particle-wall interactions.
\ifx\grdraft\undefined  \else
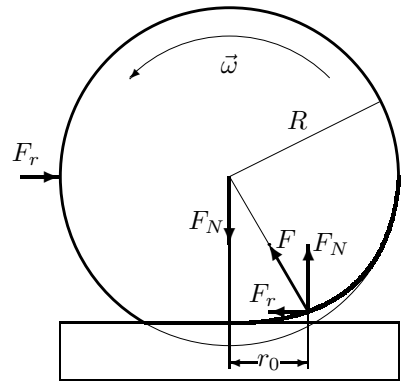
\begin{figure}[thb]
\centering
\setlength{\unitlength}{1.5mm}
\begin{picture}(35,40)(-17.5,-22.5)
\Thicklines
\put(0,0){\arc{30}{2.094}{0}}
\qbezier(14.9,0)(14,-12.99)(0,-12.99)
\put(-7.5,-12.99){\line(1,0){8}}
\thinlines
\put(0,0){\arc{30}{0}{2.094}}
\put(-8.84,8.70){\vector(-1,-1){0}}
\put(0,0){\arc{25}{-2.356}{-0.78}}
\put(0,10){\makebox(0,0){$\vec{\omega}$}}
\put(0,0){\line(2,1){13.3}}
\put(6,5){\makebox(0,0){$R$}}
\Thicklines
\put(0,0){\vector(0,-1){6}}
\put(7,-12){\vector(0,1){6}}
\put(7,-12){\vector(-1,0){3.5}}
\put(-18.5,0){\vector(1,0){3.5}}
\put(7,-12){\line(-7,12){3.3}}
\put(3.5,-6){\vector(-2,3){0}}
\thinlines
\put(5,-5.5){\makebox(0,0){$F$}}
\put(-18,2){\makebox(0,0){$F_r$}}
\put(3,-10.5){\makebox(0,0){$F_r$}}
\put(-2,-4){\makebox(0,0){$F_N$}}
\put(9,-6){\makebox(0,0){$F_N$}}
\put(7,-12){\line(-7,12){7}}
\put(7,-12){\line(0,-1){5}}
\put(0,0){\line(0,-1){17}}
\Thicklines
\thinlines
\put(4.5,-16.5){\vector(1,0){2.5}}
\put(2.5,-16.5){\vector(-1,0){2.5}}
\put(3.5,-16.5){\makebox(0,0){$r_0$}}
\thinlines
\put(-15,-17.99){\framebox(30,5)[b]{}}
\end{picture}
\caption{Viscoelastic rolling sphere on a hard surface.}
\label{sphere}
\end{figure}
\fi
%
The rolling resistance acts as a net torque  constructed out of a force couple
$F_r$ with
\begin{equation}
  F_r = |F_N| \frac{r_0}{R} (\hat{n}\times\hat{s}) .
\end{equation}
We also have to consider that the rolling resistance can only decrease angular
momentum, but never revert it. And so we have to limit $F_r$ by the quantity
that would reduce the angular momentum to zero within the next time step,
namely
\begin{equation}
  F_{r_{max}}=\frac{2}{5}m R \cdot ((\hat{n}\times\hat{s}) \cdot 
  \vec{\omega})/(\Delta t) .
\end{equation}
For the torque we therefore get
\begin{equation}
\tau = R\cdot\min(F_r,F_{r_{max}}) .
\end{equation}

As can already be seen from the definition of the rolling resistance,
Eq.~(\ref{rolling}), the rolling friction on small particles will be higher
than on large particles which was also  observed  by studying one particle on a
bumpy line~\cite{ristow94d} and in experiments with glass marbles. To
illustrate this fact, we show in Fig.~\ref{theta_rolling} the dependency of the
angle of repose on the rolling friction parameter $r_0$. The higher friction of
smaller particles results in a steeper slope of the angle of repose in
Fig.~\ref{theta_rolling}. An important result of this is that we can adjust the
rolling friction in such a way that small and large glass beads have the same
angle of repose as is seen in  experiments for glass
beads~\cite{zik94,dury97d}. Also one clearly sees that for small $r_0$ the
slope of the angle of repose $\frac{\delta \Theta}{\delta r_0}$ is proportional
to $r_0$, which result from the law of rolling resistance
Eq.~(\ref{rolling}).\\
%
\ifx\grdraft\undefined  \else
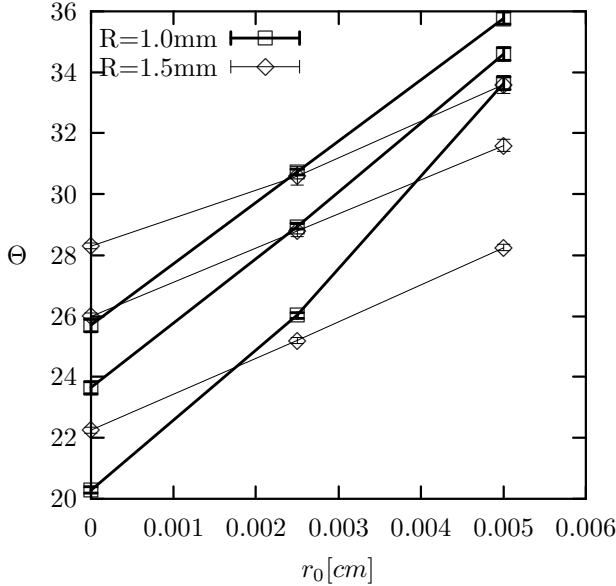
\begin{figure}
\noindent	
{\centering 
\begin{picture}(1500,850)(0,50)
\tenrm
\thicklines \path(177,179)(197,179)
\thicklines \path(954,179)(934,179)
\put(155,179){\makebox(0,0)[r]{20}}
\thicklines \path(177,275)(197,275)
\thicklines \path(954,275)(934,275)
\put(155,275){\makebox(0,0)[r]{22}}
\thicklines \path(177,371)(197,371)
\thicklines \path(954,371)(934,371)
\put(155,371){\makebox(0,0)[r]{24}}
\thicklines \path(177,466)(197,466)
\thicklines \path(954,466)(934,466)
\put(155,466){\makebox(0,0)[r]{26}}
\thicklines \path(177,562)(197,562)
\thicklines \path(954,562)(934,562)
\put(155,562){\makebox(0,0)[r]{28}}
\thicklines \path(177,658)(197,658)
\thicklines \path(954,658)(934,658)
\put(155,658){\makebox(0,0)[r]{30}}
\thicklines \path(177,754)(197,754)
\thicklines \path(954,754)(934,754)
\put(155,754){\makebox(0,0)[r]{32}}
\thicklines \path(177,849)(197,849)
\thicklines \path(954,849)(934,849)
\put(155,849){\makebox(0,0)[r]{34}}
\thicklines \path(177,945)(197,945)
\thicklines \path(954,945)(934,945)
\put(155,945){\makebox(0,0)[r]{36}}
\thicklines \path(177,179)(177,199)
\thicklines \path(177,945)(177,925)
\put(177,134){\makebox(0,0){0}}
\thicklines \path(307,179)(307,199)
\thicklines \path(307,945)(307,925)
\put(307,134){\makebox(0,0){0.001}}
\thicklines \path(436,179)(436,199)
\thicklines \path(436,945)(436,925)
\put(436,134){\makebox(0,0){0.002}}
\thicklines \path(566,179)(566,199)
\thicklines \path(566,945)(566,925)
\put(566,134){\makebox(0,0){0.003}}
\thicklines \path(695,179)(695,199)
\thicklines \path(695,945)(695,925)
\put(695,134){\makebox(0,0){0.004}}
\thicklines \path(825,179)(825,199)
\thicklines \path(825,945)(825,925)
\put(825,134){\makebox(0,0){0.005}}
\thicklines \path(954,179)(954,199)
\thicklines \path(954,945)(954,925)
\put(954,134){\makebox(0,0){0.006}}
\thicklines \path(177,179)(954,179)(954,945)(177,945)(177,179)
\put(45,562){\makebox(0,0)[l]{\shortstack{$\Theta$}}}
\put(565,67){\makebox(0,0){$r_0[cm]$}}
\put(375,903){\makebox(0,0)[r]{R=1.0mm}}
\Thicklines \path(397,903)(505,903)
\Thicklines \path(397,913)(397,893)
\Thicklines \path(505,913)(505,893)
\Thicklines \path(177,188)(177,198)
\Thicklines \path(167,188)(187,188)
\Thicklines \path(167,198)(187,198)
\Thicklines \path(501,463)(501,472)
\Thicklines \path(491,463)(511,463)
\Thicklines \path(491,472)(511,472)
\Thicklines \path(825,822)(825,842)
\Thicklines \path(815,822)(835,822)
\Thicklines \path(815,842)(835,842)
\Thicklines \path(177,344)(177,363)
\Thicklines \path(167,344)(187,344)
\Thicklines \path(167,363)(187,363)
\Thicklines \path(501,603)(501,612)
\Thicklines \path(491,603)(511,603)
\Thicklines \path(491,612)(511,612)
\Thicklines \path(825,868)(825,888)
\Thicklines \path(815,868)(835,868)
\Thicklines \path(815,888)(835,888)
\Thicklines \path(177,442)(177,461)
\Thicklines \path(167,442)(187,442)
\Thicklines \path(167,461)(187,461)
\Thicklines \path(501,688)(501,697)
\Thicklines \path(491,688)(511,688)
\Thicklines \path(491,697)(511,697)
\Thicklines \path(825,925)(825,944)
\Thicklines \path(815,925)(835,925)
\Thicklines \path(815,944)(835,944)
\put(177,193){\raisebox{-1.2pt}{\makebox(0,0){$\Box$}}}
\put(501,468){\raisebox{-1.2pt}{\makebox(0,0){$\Box$}}}
\put(825,832){\raisebox{-1.2pt}{\makebox(0,0){$\Box$}}}
\put(177,354){\raisebox{-1.2pt}{\makebox(0,0){$\Box$}}}
\put(501,607){\raisebox{-1.2pt}{\makebox(0,0){$\Box$}}}
\put(825,878){\raisebox{-1.2pt}{\makebox(0,0){$\Box$}}}
\put(177,452){\raisebox{-1.2pt}{\makebox(0,0){$\Box$}}}
\put(501,693){\raisebox{-1.2pt}{\makebox(0,0){$\Box$}}}
\put(825,934){\raisebox{-1.2pt}{\makebox(0,0){$\Box$}}}
\put(451,903){\raisebox{-1.2pt}{\makebox(0,0){$\Box$}}}
\Thicklines \path(177,193)(177,193)(501,468)(825,832)
\Thicklines \path(177,354)(177,354)(501,607)(825,878)
\Thicklines \path(177,452)(177,452)(501,693)(825,934)
\put(375,858){\makebox(0,0)[r]{R=1.5mm}}
\thinlines \path(397,858)(505,858)
\thinlines \path(397,868)(397,848)
\thinlines \path(505,868)(505,848)
\thinlines \path(177,282)(177,292)
\thinlines \path(167,282)(187,282)
\thinlines \path(167,292)(187,292)
\thinlines \path(501,423)(501,433)
\thinlines \path(491,423)(511,423)
\thinlines \path(491,433)(511,433)
\thinlines \path(825,569)(825,579)
\thinlines \path(815,569)(835,569)
\thinlines \path(815,579)(835,579)
\thinlines \path(177,461)(177,471)
\thinlines \path(167,461)(187,461)
\thinlines \path(167,471)(187,471)
\thinlines \path(501,591)(501,610)
\thinlines \path(491,591)(511,591)
\thinlines \path(491,610)(511,610)
\thinlines \path(825,725)(825,744)
\thinlines \path(815,725)(835,725)
\thinlines \path(815,744)(835,744)
\thinlines \path(177,572)(177,581)
\thinlines \path(167,572)(187,572)
\thinlines \path(167,581)(187,581)
\thinlines \path(501,672)(501,701)
\thinlines \path(491,672)(511,672)
\thinlines \path(491,701)(511,701)
\thinlines \path(825,816)(825,844)
\thinlines \path(815,816)(835,816)
\thinlines \path(815,844)(835,844)
\put(177,287){\raisebox{-1.2pt}{\makebox(0,0){$\Diamond$}}}
\put(501,428){\raisebox{-1.2pt}{\makebox(0,0){$\Diamond$}}}
\put(825,574){\raisebox{-1.2pt}{\makebox(0,0){$\Diamond$}}}
\put(177,466){\raisebox{-1.2pt}{\makebox(0,0){$\Diamond$}}}
\put(501,600){\raisebox{-1.2pt}{\makebox(0,0){$\Diamond$}}}
\put(825,734){\raisebox{-1.2pt}{\makebox(0,0){$\Diamond$}}}
\put(177,576){\raisebox{-1.2pt}{\makebox(0,0){$\Diamond$}}}
\put(501,686){\raisebox{-1.2pt}{\makebox(0,0){$\Diamond$}}}
\put(825,830){\raisebox{-1.2pt}{\makebox(0,0){$\Diamond$}}}
\put(451,858){\raisebox{-1.2pt}{\makebox(0,0){$\Diamond$}}}
\thinlines \path(177,287)(177,287)(501,428)(825,574)
\thinlines \path(177,466)(177,466)(501,600)(825,734)
\thinlines \path(177,576)(177,576)(501,686)(825,830)
\end{picture} }
 \caption{Angle of repose for different friction coefficients and 
	particle diameter of 1.0mm and 1.5mm for three different values of $\mu$ 
($\mu=0.1$, $0.2$ and $0.3$ from bottom to top).}
\label{theta_rolling}
\end{figure}
\fi
\vspace*{-1.1cm}
%

\ifx\grdraft\undefined
\newpage
\subsection*{Figure Captions}
\begin{description}
\item[\bf Figure~\ref{cylinder}:] Cross section through a more than half-filled 
      cylinder.
\item[\bf Figure~\ref{density}:] Volume fraction of small and large particles 
      in each cylinder for a volume filling fraction of 66.4\% 
\item[\bf Figure~\ref{q_time_series}:] Typical time series of the order 
      parameter $q$ for two different filling fractions of the cylinder with
      $1.0$mm and $1.5$mm beads.
\item[\bf Figure~\ref{q_fourier}:] Fourier transform of $q(t)$ of 
      Fig.\ref{q_time_series}.
\item[\bf Figure~\ref{q_max}:] Final amount of segregation for a concentration 
      of 50\% of small particles and three different size ratios.
\item[\bf Figure~\ref{q_time}:] Characteristic number of revolutions for a 
      concentration of 50\% of small particles.
\item[\bf Figure~\ref{critical}:] Final amount of segregation as function of the 
      particle size ratio $\Phi=\frac{r}{R}$  (the vertical dashed line denotes $\Phi_T$). 
\item[\bf Figure~\ref{split}:] Different snapshots of the cylinder with a 
      starting condition, where all the small (large) particles are on the right 
      (left) side of the cylinder (particle radii $1.0$mm and $1.5$mm). 
\item[\bf Figure~\ref{centroid}:] Normalized centroid position projected onto 
      the free surface of a half--filled cylinder with 1.0mm and 1.5mm 
      particles.
\item[\bf Figure~\ref{center}:] Distance of the centers of a half--filled 
      cylinder with 1.0mm and 1.5mm particles ( - - - clockwise, ----- 
      counter-clockwise rotation).
\item[\bf Figure~\ref{sphere}:] Viscoelastic rolling sphere on a hard surface. 
\item[\bf Figure~\ref{theta_rolling}:] Angle of repose for different friction 
      coefficients and particle diameter of $1.0$mm and $1.5$mm for three 
      different values of $\mu$ ($\mu=0.1$, $0.2$ and $0.3$ from bottom to top).
\end{description}
\clearpage

\begin{figure}[htb]
\centering
\setlength{\unitlength}{3.0mm}
\begin{picture}(35,35)(-17.5,-17.5)
\put(0,0){\circle{35}}
\put(0,0){\arc{30}{-0.623}{3.764}}
\put(0,0){\shade\arc{25}{-0.775}{3.917}}
\put(0,0){\arc{20}{-1.065}{4.207}}
\put(0,0){\arc{20}{-1.065}{4.207}}
\put(0,0){\circle{15}}
\put(0,0){\circle{10}}
\put(0,0){\circle{5}}
\put(0,0){\circle*{0.1}}
\put(-15,8.75){\line(1,0){30}}
\end{picture}
\caption{}
\label{cylinder}
\end{figure}
\newpage
\clearpage
\setlength{\unitlength}{0.481800pt}
\begin{figure}[htb]
{\centering 
\begin{picture}(1500,900)(0,50)
\tenrm
\thicklines \path(199,179)(219,179)
\thicklines \path(954,179)(934,179)
\put(177,179){\makebox(0,0)[r]{0}}
\thicklines \path(199,288)(219,288)
\thicklines \path(954,288)(934,288)
\put(177,288){\makebox(0,0)[r]{0.1}}
\thicklines \path(199,398)(219,398)
\thicklines \path(954,398)(934,398)
\put(177,398){\makebox(0,0)[r]{0.2}}
\thicklines \path(199,507)(219,507)
\thicklines \path(954,507)(934,507)
\put(177,507){\makebox(0,0)[r]{0.3}}
\thicklines \path(199,617)(219,617)
\thicklines \path(954,617)(934,617)
\put(177,617){\makebox(0,0)[r]{0.4}}
\thicklines \path(199,726)(219,726)
\thicklines \path(954,726)(934,726)
\put(177,726){\makebox(0,0)[r]{0.5}}
\thicklines \path(199,836)(219,836)
\thicklines \path(954,836)(934,836)
\put(177,836){\makebox(0,0)[r]{0.6}}
\thicklines \path(199,945)(219,945)
\thicklines \path(954,945)(934,945)
\put(177,945){\makebox(0,0)[r]{0.7}}
\thicklines \path(199,179)(199,199)
\thicklines \path(199,945)(199,925)
\put(199,134){\makebox(0,0){0}}
\thicklines \path(307,179)(307,199)
\thicklines \path(307,945)(307,925)
\put(307,134){\makebox(0,0){0.5}}
\thicklines \path(415,179)(415,199)
\thicklines \path(415,945)(415,925)
\put(415,134){\makebox(0,0){1}}
\thicklines \path(523,179)(523,199)
\thicklines \path(523,945)(523,925)
\put(523,134){\makebox(0,0){1.5}}
\thicklines \path(630,179)(630,199)
\thicklines \path(630,945)(630,925)
\put(630,134){\makebox(0,0){2}}
\thicklines \path(738,179)(738,199)
\thicklines \path(738,945)(738,925)
\put(738,134){\makebox(0,0){2.5}}
\thicklines \path(846,179)(846,199)
\thicklines \path(846,945)(846,925)
\put(846,134){\makebox(0,0){3}}
\thicklines \path(954,179)(954,199)
\thicklines \path(954,945)(954,925)
\put(954,134){\makebox(0,0){3.5}}
\thicklines \path(199,179)(954,179)(954,945)(199,945)(199,179)
\put(45,562){\makebox(0,0)[l]{\shortstack{$\eta$}}}
\put(576,67){\makebox(0,0){distance[cm]}}
\put(780,312){\makebox(0,0)[r]{small}}
\thinlines \path(802,312)(910,312)
\thinlines \path(230,768)(230,768)(293,743)(356,676)(419,668)(482,683)(545,671)(608,642)(671,624)(734,554)(797,451)(860,345)
\put(230,768){\raisebox{-1.2pt}{\makebox(0,0){$\Diamond$}}}
\put(293,743){\raisebox{-1.2pt}{\makebox(0,0){$\Diamond$}}}
\put(356,676){\raisebox{-1.2pt}{\makebox(0,0){$\Diamond$}}}
\put(419,668){\raisebox{-1.2pt}{\makebox(0,0){$\Diamond$}}}
\put(482,683){\raisebox{-1.2pt}{\makebox(0,0){$\Diamond$}}}
\put(545,671){\raisebox{-1.2pt}{\makebox(0,0){$\Diamond$}}}
\put(608,642){\raisebox{-1.2pt}{\makebox(0,0){$\Diamond$}}}
\put(671,624){\raisebox{-1.2pt}{\makebox(0,0){$\Diamond$}}}
\put(734,554){\raisebox{-1.2pt}{\makebox(0,0){$\Diamond$}}}
\put(797,451){\raisebox{-1.2pt}{\makebox(0,0){$\Diamond$}}}
\put(860,345){\raisebox{-1.2pt}{\makebox(0,0){$\Diamond$}}}
\put(856,312){\raisebox{-1.2pt}{\makebox(0,0){$\Diamond$}}}
\put(780,267){\makebox(0,0)[r]{large}}
\thicklines \path(802,267)(910,267)
\thicklines \path(230,233)(230,233)(293,272)(356,341)(419,400)(482,365)(545,374)(608,412)(671,424)(734,502)(797,610)(860,712)
\put(230,233){\makebox(0,0){$+$}}
\put(293,272){\makebox(0,0){$+$}}
\put(356,341){\makebox(0,0){$+$}}
\put(419,400){\makebox(0,0){$+$}}
\put(482,365){\makebox(0,0){$+$}}
\put(545,374){\makebox(0,0){$+$}}
\put(608,412){\makebox(0,0){$+$}}
\put(671,424){\makebox(0,0){$+$}}
\put(734,502){\makebox(0,0){$+$}}
\put(797,610){\makebox(0,0){$+$}}
\put(860,712){\makebox(0,0){$+$}}
\put(856,267){\makebox(0,0){$+$}}
\put(780,222){\makebox(0,0)[r]{total}}
\Thicklines \path(802,222)(910,222)
\Thicklines \path(230,822)(230,822)(293,836)(356,838)(419,889)(482,869)(545,866)(608,875)(671,869)(734,877)(797,882)(860,878)
\put(230,822){\raisebox{-1.2pt}{\makebox(0,0){$\Box$}}}
\put(293,836){\raisebox{-1.2pt}{\makebox(0,0){$\Box$}}}
\put(356,838){\raisebox{-1.2pt}{\makebox(0,0){$\Box$}}}
\put(419,889){\raisebox{-1.2pt}{\makebox(0,0){$\Box$}}}
\put(482,869){\raisebox{-1.2pt}{\makebox(0,0){$\Box$}}}
\put(545,866){\raisebox{-1.2pt}{\makebox(0,0){$\Box$}}}
\put(608,875){\raisebox{-1.2pt}{\makebox(0,0){$\Box$}}}
\put(671,869){\raisebox{-1.2pt}{\makebox(0,0){$\Box$}}}
\put(734,877){\raisebox{-1.2pt}{\makebox(0,0){$\Box$}}}
\put(797,882){\raisebox{-1.2pt}{\makebox(0,0){$\Box$}}}
\put(860,878){\raisebox{-1.2pt}{\makebox(0,0){$\Box$}}}
\put(856,222){\raisebox{-1.2pt}{\makebox(0,0){$\Box$}}}
\end{picture} }
 \caption{}
\label{density}
\end{figure}
\newpage
\clearpage

\begin{figure}[htb]
{\centering 
\begin{picture}(1500,900)(0,50)
\tenrm
\thicklines \path(199,179)(219,179)
\thicklines \path(954,179)(934,179)
\put(177,179){\makebox(0,0)[r]{0}}
\thicklines \path(199,332)(219,332)
\thicklines \path(954,332)(934,332)
\put(177,332){\makebox(0,0)[r]{0.2}}
\thicklines \path(199,485)(219,485)
\thicklines \path(954,485)(934,485)
\put(177,485){\makebox(0,0)[r]{0.4}}
\thicklines \path(199,639)(219,639)
\thicklines \path(954,639)(934,639)
\put(177,639){\makebox(0,0)[r]{0.6}}
\thicklines \path(199,792)(219,792)
\thicklines \path(954,792)(934,792)
\put(177,792){\makebox(0,0)[r]{0.8}}
\thicklines \path(199,945)(219,945)
\thicklines \path(954,945)(934,945)
\put(177,945){\makebox(0,0)[r]{1}}
\thicklines \path(199,179)(199,199)
\thicklines \path(199,945)(199,925)
\put(199,134){\makebox(0,0){0}}
\thicklines \path(307,179)(307,199)
\thicklines \path(307,945)(307,925)
\put(307,134){\makebox(0,0){5}}
\thicklines \path(415,179)(415,199)
\thicklines \path(415,945)(415,925)
\put(415,134){\makebox(0,0){10}}
\thicklines \path(523,179)(523,199)
\thicklines \path(523,945)(523,925)
\put(523,134){\makebox(0,0){15}}
\thicklines \path(630,179)(630,199)
\thicklines \path(630,945)(630,925)
\put(630,134){\makebox(0,0){20}}
\thicklines \path(738,179)(738,199)
\thicklines \path(738,945)(738,925)
\put(738,134){\makebox(0,0){25}}
\thicklines \path(846,179)(846,199)
\thicklines \path(846,945)(846,925)
\put(846,134){\makebox(0,0){30}}
\thicklines \path(954,179)(954,199)
\thicklines \path(954,945)(954,925)
\put(954,134){\makebox(0,0){35}}
\thicklines \path(199,179)(954,179)(954,945)(199,945)(199,179)
\put(45,562){\makebox(0,0)[l]{\shortstack{q(t)}}}
\put(576,67){\makebox(0,0){t[s]}}
\put(485,903){\makebox(0,0)[r]{50\% filled}}
\thinlines \path(507,903)(615,903)
\thinlines \path(231,252)(231,252)(235,252)(240,262)(240,273)(244,254)(249,233)(251,268)(256,286)(260,242)(262,253)(267,313)(271,296)(273,338)(278,394)(282,426)(284,451)(289,497)(293,426)(295,386)(300,381)(304,404)(306,491)(311,381)(315,374)(318,385)(322,423)(326,403)(329,475)(333,495)(338,603)(340,649)(345,521)(349,490)(351,512)(356,546)(360,518)(363,523)(367,475)(372,451)(374,503)(378,463)(383,517)(385,584)(390,580)(394,688)(396,692)(401,618)(405,563)(408,572)(412,625)
\thinlines \path(412,625)(416,534)(419,558)(423,506)(426,545)(431,551)(435,570)(437,545)(442,574)(446,694)(448,718)(453,705)(457,602)(460,591)(464,620)(468,585)(471,596)(475,571)(479,554)(482,564)(486,562)(490,594)(493,622)(497,626)(501,716)(504,771)(508,755)(512,692)(515,663)(519,619)(524,627)(528,597)(530,613)(535,597)(539,597)(541,631)(545,691)(550,664)(552,719)(556,665)(561,752)(563,694)(567,738)(572,679)(574,674)(578,690)(583,642)(585,629)(589,613)(594,629)(596,620)
\thinlines \path(596,620)(600,674)(605,682)(607,709)(611,734)(616,772)(618,766)(622,711)(627,683)(629,742)(633,655)(638,645)(640,628)(644,642)(649,651)(651,679)(655,688)(660,742)(662,762)(666,710)(671,745)(673,739)(677,686)(682,724)(684,760)(688,678)(693,640)(695,633)(699,615)(704,643)(706,694)(710,746)(715,776)(717,783)(722,753)(726,696)(728,684)(733,698)(737,663)(739,731)(744,704)(748,640)(750,663)(755,648)(759,684)(761,682)(766,705)(770,805)(771,782)(776,746)(780,712)
\thinlines \path(780,712)(782,714)(787,729)(791,689)(795,673)(798,676)(802,680)(806,662)(809,653)(813,678)(817,662)(820,687)(824,734)(828,732)(831,811)(835,701)(840,685)(844,739)(846,712)(851,705)(855,701)(857,703)(862,680)(866,680)(868,662)(873,687)(877,751)(880,774)(884,764)(888,719)(891,686)(895,710)(900,656)(902,642)(906,661)(911,721)(913,713)(917,688)(922,665)(924,720)(928,716)(933,777)(935,743)(939,739)(944,693)(946,693)(950,717)(954,676)
\put(485,858){\makebox(0,0)[r]{80\% filled}}
\thicklines \path(507,858)(615,858)
\thicklines \path(231,252)(231,252)(235,252)(236,252)(241,254)(242,254)(246,250)(248,249)(252,250)(253,249)(258,258)(259,254)(263,260)(265,258)(269,251)(271,252)(275,264)(276,259)(281,267)(282,273)(287,286)(288,284)(293,306)(294,310)(298,311)(300,318)(304,339)(306,339)(310,336)(311,331)(316,323)(317,325)(322,336)(323,334)(327,344)(331,343)(333,343)(337,342)(339,344)(343,353)(344,353)(349,362)(350,364)(355,375)(356,378)(360,369)(361,379)(366,383)(367,382)(372,405)(373,408)
\thicklines \path(373,408)(377,422)(378,419)(383,406)(384,403)(388,402)(389,406)(394,406)(395,406)(400,404)(401,407)(405,414)(406,419)(411,420)(412,417)(416,421)(418,420)(422,437)(423,440)(428,432)(429,432)(433,440)(434,440)(439,442)(440,449)(444,474)(446,475)(450,470)(451,471)(456,456)(457,454)(461,459)(462,461)(467,459)(468,455)(473,459)(474,459)(478,468)(479,476)(484,477)(485,473)(489,480)(490,481)(495,493)(496,493)(501,490)(502,490)(506,497)(506,497)(511,511)(512,516)
\thicklines \path(512,516)(516,515)(517,516)(522,510)(523,505)(528,504)(529,505)(533,512)(534,508)(539,506)(540,507)(544,520)(545,520)(550,524)(551,521)(556,518)(557,519)(561,526)(562,527)(567,536)(568,537)(572,526)(573,525)(578,540)(579,539)(584,553)(585,550)(589,543)(590,545)(595,540)(596,537)(600,540)(601,542)(606,546)(607,541)
\thinlines \path(239,179)(245,205)(252,236)(260,266)(268,294)(275,320)(283,344)(291,367)(298,389)(306,409)(313,428)(321,445)(329,462)(336,477)(344,492)(352,505)(359,518)(367,530)(374,541)(382,552)(390,561)(397,571)(405,579)(413,587)(420,595)(428,602)(435,609)(443,615)(451,621)(458,626)(466,631)(474,636)(481,641)(489,645)(496,649)(504,652)(512,656)(519,659)(527,662)(535,665)(542,668)(550,670)(557,673)(565,675)(573,677)(580,679)(588,681)(596,682)(603,684)(611,685)(618,687)
\thinlines \path(618,687)(626,688)(634,689)(641,690)(649,692)(657,693)(664,693)(672,694)(679,695)(687,696)(695,697)(702,697)(710,698)(718,699)(725,699)(733,700)(740,700)(748,701)(756,701)(763,702)(771,702)(779,702)(786,703)(794,703)(801,703)(809,703)(817,704)(824,704)(832,704)(840,704)(847,705)(855,705)(862,705)(870,705)(878,705)(885,705)(893,706)(901,706)(908,706)(916,706)(923,706)(931,706)(939,706)(946,706)(954,706)
\end{picture} }
\caption{}
\label{q_time_series}
\end{figure}
\newpage
\clearpage

%
\begin{figure}[t]
{\centering \input fourier.tex }
\caption{}
\label{q_fourier}
\end{figure}
\newpage
\clearpage

\begin{figure}[htb]
{\centering 
\begin{picture}(1500,900)(0,100)
\tenrm
\thicklines \path(199,179)(219,179)
\thicklines \path(954,179)(934,179)
\put(177,179){\makebox(0,0)[r]{0}}
\thicklines \path(199,332)(219,332)
\thicklines \path(954,332)(934,332)
\put(177,332){\makebox(0,0)[r]{0.2}}
\thicklines \path(199,485)(219,485)
\thicklines \path(954,485)(934,485)
\put(177,485){\makebox(0,0)[r]{0.4}}
\thicklines \path(199,639)(219,639)
\thicklines \path(954,639)(934,639)
\put(177,639){\makebox(0,0)[r]{0.6}}
\thicklines \path(199,792)(219,792)
\thicklines \path(954,792)(934,792)
\put(177,792){\makebox(0,0)[r]{0.8}}
\thicklines \path(199,945)(219,945)
\thicklines \path(954,945)(934,945)
\put(177,945){\makebox(0,0)[r]{1}}
\thicklines \path(199,179)(199,199)
\thicklines \path(199,945)(199,925)
\put(199,134){\makebox(0,0){0}}
\thicklines \path(350,179)(350,199)
\thicklines \path(350,945)(350,925)
\put(350,134){\makebox(0,0){20}}
\thicklines \path(501,179)(501,199)
\thicklines \path(501,945)(501,925)
\put(501,134){\makebox(0,0){40}}
\thicklines \path(652,179)(652,199)
\thicklines \path(652,945)(652,925)
\put(652,134){\makebox(0,0){60}}
\thicklines \path(803,179)(803,199)
\thicklines \path(803,945)(803,925)
\put(803,134){\makebox(0,0){80}}
\thicklines \path(954,179)(954,199)
\thicklines \path(954,945)(954,925)
\put(954,134){\makebox(0,0){100}}
\thicklines \path(199,179)(954,179)(954,945)(199,945)(199,179)
\put(45,562){\makebox(0,0)[l]{\shortstack{$q_\infty$}}}
\put(576,67){\makebox(0,0){filling fraction of the drum}}
\put(375,903){\makebox(0,0)[r]{0.75mm}}
\thinlines \path(397,903)(505,903)
\thinlines \path(267,180)(267,180)(366,330)(426,439)(493,568)(520,622)(614,801)(700,883)(801,616)
\put(267,180){\raisebox{-1.2pt}{\makebox(0,0){$\Diamond$}}}
\put(366,330){\raisebox{-1.2pt}{\makebox(0,0){$\Diamond$}}}
\put(426,439){\raisebox{-1.2pt}{\makebox(0,0){$\Diamond$}}}
\put(493,568){\raisebox{-1.2pt}{\makebox(0,0){$\Diamond$}}}
\put(520,622){\raisebox{-1.2pt}{\makebox(0,0){$\Diamond$}}}
\put(614,801){\raisebox{-1.2pt}{\makebox(0,0){$\Diamond$}}}
\put(700,883){\raisebox{-1.2pt}{\makebox(0,0){$\Diamond$}}}
\put(801,616){\raisebox{-1.2pt}{\makebox(0,0){$\Diamond$}}}
\put(451,903){\raisebox{-1.2pt}{\makebox(0,0){$\Diamond$}}}
\thinlines \path(267,179)(267,195)
\thinlines \path(257,179)(277,179)
\thinlines \path(257,195)(277,195)
\thinlines \path(366,307)(366,353)
\thinlines \path(356,307)(376,307)
\thinlines \path(356,353)(376,353)
\thinlines \path(426,416)(426,462)
\thinlines \path(416,416)(436,416)
\thinlines \path(416,462)(436,462)
\thinlines \path(493,545)(493,591)
\thinlines \path(483,545)(503,545)
\thinlines \path(483,591)(503,591)
\thinlines \path(520,583)(520,660)
\thinlines \path(510,583)(530,583)
\thinlines \path(510,660)(530,660)
\thinlines \path(614,732)(614,870)
\thinlines \path(604,732)(624,732)
\thinlines \path(604,870)(624,870)
\thinlines \path(700,806)(700,945)
\thinlines \path(690,806)(710,806)
\thinlines \path(690,945)(710,945)
\thinlines \path(801,559)(801,674)
\thinlines \path(791,559)(811,559)
\thinlines \path(791,674)(811,674)
\put(267,180){\raisebox{-1.2pt}{\makebox(0,0){$\Diamond$}}}
\put(366,330){\raisebox{-1.2pt}{\makebox(0,0){$\Diamond$}}}
\put(426,439){\raisebox{-1.2pt}{\makebox(0,0){$\Diamond$}}}
\put(493,568){\raisebox{-1.2pt}{\makebox(0,0){$\Diamond$}}}
\put(520,622){\raisebox{-1.2pt}{\makebox(0,0){$\Diamond$}}}
\put(614,801){\raisebox{-1.2pt}{\makebox(0,0){$\Diamond$}}}
\put(700,883){\raisebox{-1.2pt}{\makebox(0,0){$\Diamond$}}}
\put(801,616){\raisebox{-1.2pt}{\makebox(0,0){$\Diamond$}}}
\put(375,858){\makebox(0,0)[r]{1.0mm}}
\thicklines \path(397,858)(505,858)
\thicklines \path(253,181)(253,181)(347,356)(458,570)(577,672)(695,740)(806,610)(900,362)
\put(253,181){\makebox(0,0){$+$}}
\put(347,356){\makebox(0,0){$+$}}
\put(458,570){\makebox(0,0){$+$}}
\put(577,672){\makebox(0,0){$+$}}
\put(695,740){\makebox(0,0){$+$}}
\put(806,610){\makebox(0,0){$+$}}
\put(900,362){\makebox(0,0){$+$}}
\put(451,858){\makebox(0,0){$+$}}
\thicklines \path(253,179)(253,212)
\thicklines \path(243,179)(263,179)
\thicklines \path(243,212)(263,212)
\thicklines \path(347,325)(347,387)
\thicklines \path(337,325)(357,325)
\thicklines \path(337,387)(357,387)
\thicklines \path(458,539)(458,600)
\thicklines \path(448,539)(468,539)
\thicklines \path(448,600)(468,600)
\thicklines \path(577,642)(577,703)
\thicklines \path(567,642)(587,642)
\thicklines \path(567,703)(587,703)
\thicklines \path(695,705)(695,774)
\thicklines \path(685,705)(705,705)
\thicklines \path(685,774)(705,774)
\thicklines \path(806,587)(806,633)
\thicklines \path(796,587)(816,587)
\thicklines \path(796,633)(816,633)
\thicklines \path(900,347)(900,377)
\thicklines \path(890,347)(910,347)
\thicklines \path(890,377)(910,377)
\put(253,181){\makebox(0,0){$+$}}
\put(347,356){\makebox(0,0){$+$}}
\put(458,570){\makebox(0,0){$+$}}
\put(577,672){\makebox(0,0){$+$}}
\put(695,740){\makebox(0,0){$+$}}
\put(806,610){\makebox(0,0){$+$}}
\put(900,362){\makebox(0,0){$+$}}
\put(375,813){\makebox(0,0)[r]{1.25mm}}
\Thicklines \path(397,813)(505,813)
\Thicklines \path(253,194)(253,194)(347,249)(458,320)(577,397)(695,559)(806,505)(900,346)
\put(253,194){\raisebox{-1.2pt}{\makebox(0,0){$\Box$}}}
\put(347,249){\raisebox{-1.2pt}{\makebox(0,0){$\Box$}}}
\put(458,320){\raisebox{-1.2pt}{\makebox(0,0){$\Box$}}}
\put(577,397){\raisebox{-1.2pt}{\makebox(0,0){$\Box$}}}
\put(695,559){\raisebox{-1.2pt}{\makebox(0,0){$\Box$}}}
\put(806,505){\raisebox{-1.2pt}{\makebox(0,0){$\Box$}}}
\put(900,346){\raisebox{-1.2pt}{\makebox(0,0){$\Box$}}}
\put(451,813){\raisebox{-1.2pt}{\makebox(0,0){$\Box$}}}
\Thicklines \path(253,179)(253,225)
\Thicklines \path(243,179)(263,179)
\Thicklines \path(243,225)(263,225)
\Thicklines \path(347,234)(347,265)
\Thicklines \path(337,234)(357,234)
\Thicklines \path(337,265)(357,265)
\Thicklines \path(458,297)(458,343)
\Thicklines \path(448,297)(468,297)
\Thicklines \path(448,343)(468,343)
\Thicklines \path(577,367)(577,428)
\Thicklines \path(567,367)(587,367)
\Thicklines \path(567,428)(587,428)
\Thicklines \path(695,513)(695,605)
\Thicklines \path(685,513)(705,513)
\Thicklines \path(685,605)(705,605)
\Thicklines \path(806,429)(806,582)
\Thicklines \path(796,429)(816,429)
\Thicklines \path(796,582)(816,582)
\Thicklines \path(900,323)(900,369)
\Thicklines \path(890,323)(910,323)
\Thicklines \path(890,369)(910,369)
\put(253,194){\raisebox{-1.2pt}{\makebox(0,0){$\Box$}}}
\put(347,249){\raisebox{-1.2pt}{\makebox(0,0){$\Box$}}}
\put(458,320){\raisebox{-1.2pt}{\makebox(0,0){$\Box$}}}
\put(577,397){\raisebox{-1.2pt}{\makebox(0,0){$\Box$}}}
\put(695,559){\raisebox{-1.2pt}{\makebox(0,0){$\Box$}}}
\put(806,505){\raisebox{-1.2pt}{\makebox(0,0){$\Box$}}}
\put(900,346){\raisebox{-1.2pt}{\makebox(0,0){$\Box$}}}
\end{picture} }
 \caption{}
\label{q_max}
\end{figure}
\newpage
\clearpage

\begin{figure}[htb]
{\centering 
\begin{picture}(1500,900)(0,100)
\tenrm
\thicklines \path(177,179)(197,179)
\thicklines \path(954,179)(934,179)
\put(155,179){\makebox(0,0)[r]{0}}
\thicklines \path(177,332)(197,332)
\thicklines \path(954,332)(934,332)
\put(155,332){\makebox(0,0)[r]{2}}
\thicklines \path(177,485)(197,485)
\thicklines \path(954,485)(934,485)
\put(155,485){\makebox(0,0)[r]{4}}
\thicklines \path(177,639)(197,639)
\thicklines \path(954,639)(934,639)
\put(155,639){\makebox(0,0)[r]{6}}
\thicklines \path(177,792)(197,792)
\thicklines \path(954,792)(934,792)
\put(155,792){\makebox(0,0)[r]{8}}
\thicklines \path(177,945)(197,945)
\thicklines \path(954,945)(934,945)
\put(155,945){\makebox(0,0)[r]{10}}
\thicklines \path(177,179)(177,199)
\thicklines \path(177,945)(177,925)
\put(177,134){\makebox(0,0){0}}
\thicklines \path(332,179)(332,199)
\thicklines \path(332,945)(332,925)
\put(332,134){\makebox(0,0){20}}
\thicklines \path(488,179)(488,199)
\thicklines \path(488,945)(488,925)
\put(488,134){\makebox(0,0){40}}
\thicklines \path(643,179)(643,199)
\thicklines \path(643,945)(643,925)
\put(643,134){\makebox(0,0){60}}
\thicklines \path(799,179)(799,199)
\thicklines \path(799,945)(799,925)
\put(799,134){\makebox(0,0){80}}
\thicklines \path(954,179)(954,199)
\thicklines \path(954,945)(954,925)
\put(954,134){\makebox(0,0){100}}
\thicklines \path(177,179)(954,179)(954,945)(177,945)(177,179)
\put(45,562){\makebox(0,0)[l]{\shortstack{$n_c$}}}
\put(565,67){\makebox(0,0){filling fraction of the drum}}
\put(353,903){\makebox(0,0)[r]{0.75mm}}
\thinlines \path(375,903)(483,903)
\thinlines \path(410,236)(410,236)(479,239)(507,269)(604,309)(693,273)(797,280)
\put(410,236){\raisebox{-1.2pt}{\makebox(0,0){$\Diamond$}}}
\put(479,239){\raisebox{-1.2pt}{\makebox(0,0){$\Diamond$}}}
\put(507,269){\raisebox{-1.2pt}{\makebox(0,0){$\Diamond$}}}
\put(604,309){\raisebox{-1.2pt}{\makebox(0,0){$\Diamond$}}}
\put(693,273){\raisebox{-1.2pt}{\makebox(0,0){$\Diamond$}}}
\put(797,280){\raisebox{-1.2pt}{\makebox(0,0){$\Diamond$}}}
\put(429,903){\raisebox{-1.2pt}{\makebox(0,0){$\Diamond$}}}
\thinlines \path(410,226)(410,246)
\thinlines \path(400,226)(420,226)
\thinlines \path(400,246)(420,246)
\thinlines \path(479,236)(479,243)
\thinlines \path(469,236)(489,236)
\thinlines \path(469,243)(489,243)
\thinlines \path(507,257)(507,282)
\thinlines \path(497,257)(517,257)
\thinlines \path(497,282)(517,282)
\thinlines \path(604,292)(604,327)
\thinlines \path(594,292)(614,292)
\thinlines \path(594,327)(614,327)
\thinlines \path(693,259)(693,287)
\thinlines \path(683,259)(703,259)
\thinlines \path(683,287)(703,287)
\thinlines \path(797,259)(797,302)
\thinlines \path(787,259)(807,259)
\thinlines \path(787,302)(807,302)
\put(410,236){\raisebox{-1.2pt}{\makebox(0,0){$\Diamond$}}}
\put(479,239){\raisebox{-1.2pt}{\makebox(0,0){$\Diamond$}}}
\put(507,269){\raisebox{-1.2pt}{\makebox(0,0){$\Diamond$}}}
\put(604,309){\raisebox{-1.2pt}{\makebox(0,0){$\Diamond$}}}
\put(693,273){\raisebox{-1.2pt}{\makebox(0,0){$\Diamond$}}}
\put(797,280){\raisebox{-1.2pt}{\makebox(0,0){$\Diamond$}}}
\put(353,858){\makebox(0,0)[r]{1.0mm}}
\thicklines \path(375,858)(483,858)
\thicklines \path(329,240)(329,240)(444,232)(566,295)(687,404)(802,381)(898,353)
\put(329,240){\makebox(0,0){$+$}}
\put(444,232){\makebox(0,0){$+$}}
\put(566,295){\makebox(0,0){$+$}}
\put(687,404){\makebox(0,0){$+$}}
\put(802,381){\makebox(0,0){$+$}}
\put(898,353){\makebox(0,0){$+$}}
\put(429,858){\makebox(0,0){$+$}}
\thicklines \path(329,230)(329,251)
\thicklines \path(319,230)(339,230)
\thicklines \path(319,251)(339,251)
\thicklines \path(444,228)(444,237)
\thicklines \path(434,228)(454,228)
\thicklines \path(434,237)(454,237)
\thicklines \path(566,289)(566,301)
\thicklines \path(556,289)(576,289)
\thicklines \path(556,301)(576,301)
\thicklines \path(687,391)(687,418)
\thicklines \path(677,391)(697,391)
\thicklines \path(677,418)(697,418)
\thicklines \path(802,370)(802,391)
\thicklines \path(792,370)(812,370)
\thicklines \path(792,391)(812,391)
\thicklines \path(898,345)(898,361)
\thicklines \path(888,345)(908,345)
\thicklines \path(888,361)(908,361)
\put(329,240){\makebox(0,0){$+$}}
\put(444,232){\makebox(0,0){$+$}}
\put(566,295){\makebox(0,0){$+$}}
\put(687,404){\makebox(0,0){$+$}}
\put(802,381){\makebox(0,0){$+$}}
\put(898,353){\makebox(0,0){$+$}}
\put(353,813){\makebox(0,0)[r]{1.25mm}}
\Thicklines \path(375,813)(483,813)
\Thicklines \path(329,316)(329,316)(444,324)(566,363)(687,521)(802,817)(898,413)
\put(329,316){\raisebox{-1.2pt}{\makebox(0,0){$\Box$}}}
\put(444,324){\raisebox{-1.2pt}{\makebox(0,0){$\Box$}}}
\put(566,363){\raisebox{-1.2pt}{\makebox(0,0){$\Box$}}}
\put(687,521){\raisebox{-1.2pt}{\makebox(0,0){$\Box$}}}
\put(802,817){\raisebox{-1.2pt}{\makebox(0,0){$\Box$}}}
\put(898,413){\raisebox{-1.2pt}{\makebox(0,0){$\Box$}}}
\put(429,813){\raisebox{-1.2pt}{\makebox(0,0){$\Box$}}}
\Thicklines \path(329,296)(329,335)
\Thicklines \path(319,296)(339,296)
\Thicklines \path(319,335)(339,335)
\Thicklines \path(444,313)(444,335)
\Thicklines \path(434,313)(454,313)
\Thicklines \path(434,335)(454,335)
\Thicklines \path(566,345)(566,381)
\Thicklines \path(556,345)(576,345)
\Thicklines \path(556,381)(576,381)
\Thicklines \path(687,490)(687,551)
\Thicklines \path(677,490)(697,490)
\Thicklines \path(677,551)(697,551)
\Thicklines \path(802,711)(802,924)
\Thicklines \path(792,711)(812,711)
\Thicklines \path(792,924)(812,924)
\Thicklines \path(898,398)(898,427)
\Thicklines \path(888,398)(908,398)
\Thicklines \path(888,427)(908,427)
\put(329,316){\raisebox{-1.2pt}{\makebox(0,0){$\Box$}}}
\put(444,324){\raisebox{-1.2pt}{\makebox(0,0){$\Box$}}}
\put(566,363){\raisebox{-1.2pt}{\makebox(0,0){$\Box$}}}
\put(687,521){\raisebox{-1.2pt}{\makebox(0,0){$\Box$}}}
\put(802,817){\raisebox{-1.2pt}{\makebox(0,0){$\Box$}}}
\put(898,413){\raisebox{-1.2pt}{\makebox(0,0){$\Box$}}}
\end{picture} }
\caption{}
\label{q_time}
\end{figure}
\newpage
\clearpage
\addtolength{\textheight}{2cm}
\begin{figure}[htb]
{\centering 
\begin{picture}(1500,900)(0,50)
\tenrm
\thicklines \path(199,179)(219,179)
\thicklines \path(954,179)(934,179)
\put(177,179){\makebox(0,0)[r]{0}}
\thicklines \path(199,332)(219,332)
\thicklines \path(954,332)(934,332)
\put(177,332){\makebox(0,0)[r]{0.2}}
\thicklines \path(199,485)(219,485)
\thicklines \path(954,485)(934,485)
\put(177,485){\makebox(0,0)[r]{0.4}}
\thicklines \path(199,639)(219,639)
\thicklines \path(954,639)(934,639)
\put(177,639){\makebox(0,0)[r]{0.6}}
\thicklines \path(199,792)(219,792)
\thicklines \path(954,792)(934,792)
\put(177,792){\makebox(0,0)[r]{0.8}}
\thicklines \path(199,945)(219,945)
\thicklines \path(954,945)(934,945)
\put(177,945){\makebox(0,0)[r]{1}}
\thicklines \path(199,179)(199,199)
\thicklines \path(199,945)(199,925)
\put(199,134){\makebox(0,0){0}}
\thicklines \path(325,179)(325,199)
\thicklines \path(325,945)(325,925)
\put(325,134){\makebox(0,0){0.2}}
\dashline{10}(297,179)(297,945)
\thicklines \path(451,179)(451,199)
\thicklines \path(451,945)(451,925)
\put(451,134){\makebox(0,0){0.4}}
\thicklines \path(577,179)(577,199)
\thicklines \path(577,945)(577,925)
\put(577,134){\makebox(0,0){0.6}}
\thicklines \path(702,179)(702,199)
\thicklines \path(702,945)(702,925)
\put(702,134){\makebox(0,0){0.8}}
\thicklines \path(828,179)(828,199)
\thicklines \path(828,945)(828,925)
\put(828,134){\makebox(0,0){1}}
\thicklines \path(954,179)(954,199)
\thicklines \path(954,945)(954,925)
\put(954,134){\makebox(0,0){1.2}}
\thicklines \path(199,179)(954,179)(954,945)(199,945)(199,179)
\put(45,562){\makebox(0,0)[l]{\shortstack{$q_\infty$}}}
\put(576,67){\makebox(0,0){$\Phi$}}
\put(780,903){\makebox(0,0)[r]{data}}
\thinlines \path(802,903)(910,903)
\thinlines \path(802,913)(802,893)
\thinlines \path(910,913)(910,893)
\thinlines \path(514,691)(514,829)
\thinlines \path(504,691)(524,691)
\thinlines \path(504,829)(524,829)
\thinlines \path(618,642)(618,703)
\thinlines \path(608,642)(628,642)
\thinlines \path(608,703)(628,703)
\thinlines \path(671,434)(671,511)
\thinlines \path(661,434)(681,434)
\thinlines \path(661,511)(681,511)
\thinlines \path(723,362)(723,423)
\thinlines \path(713,362)(733,362)
\thinlines \path(713,423)(733,423)
\thinlines \path(750,287)(750,327)
\thinlines \path(740,287)(760,287)
\thinlines \path(740,327)(760,327)
\thinlines \path(776,246)(776,277)
\thinlines \path(766,246)(786,246)
\thinlines \path(766,277)(786,277)
\thinlines \path(802,217)(802,248)
\thinlines \path(792,217)(812,217)
\thinlines \path(792,248)(812,248)
\put(514,760){\raisebox{-1.2pt}{\makebox(0,0){$\diamond$}}}
\put(618,672){\raisebox{-1.2pt}{\makebox(0,0){$\diamond$}}}
\put(671,472){\raisebox{-1.2pt}{\makebox(0,0){$\diamond$}}}
\put(723,393){\raisebox{-1.2pt}{\makebox(0,0){$\diamond$}}}
\put(750,307){\raisebox{-1.2pt}{\makebox(0,0){$\diamond$}}}
\put(776,262){\raisebox{-1.2pt}{\makebox(0,0){$\diamond$}}}
\put(802,233){\raisebox{-1.2pt}{\makebox(0,0){$\diamond$}}}
\put(856,903){\raisebox{-1.2pt}{\makebox(0,0){$\diamond$}}}
\put(780,858){\makebox(0,0)[r]{fit}}
\thinlines \drawline[-50](802,858)(910,858)
\thinlines \drawline[-50](441,945)(443,941)(451,926)(458,911)(466,896)(474,881)(481,865)(489,850)(496,835)(504,820)(512,805)(519,789)(527,774)(535,759)(542,744)(550,729)(557,713)(565,698)(573,683)(580,668)(588,653)(596,637)(603,622)(611,607)(618,592)(626,577)(634,561)(641,546)(649,531)(657,516)(664,501)(672,485)(679,470)(687,455)(695,440)(702,425)(710,410)(718,394)(725,379)(733,364)(740,349)(748,334)(756,318)(763,303)(771,288)(779,273)(786,258)(794,242)(801,227)(809,212)(817,197)
\thinlines \drawline[-50](817,197)(824,182)(826,179)
\end{picture} }
 \caption{}
\label{critical}
\end{figure}
\newpage
\clearpage
\begin{figure}[htb]
\noindent	
\begin{minipage}[b]{.49\linewidth}	
\psfig{file=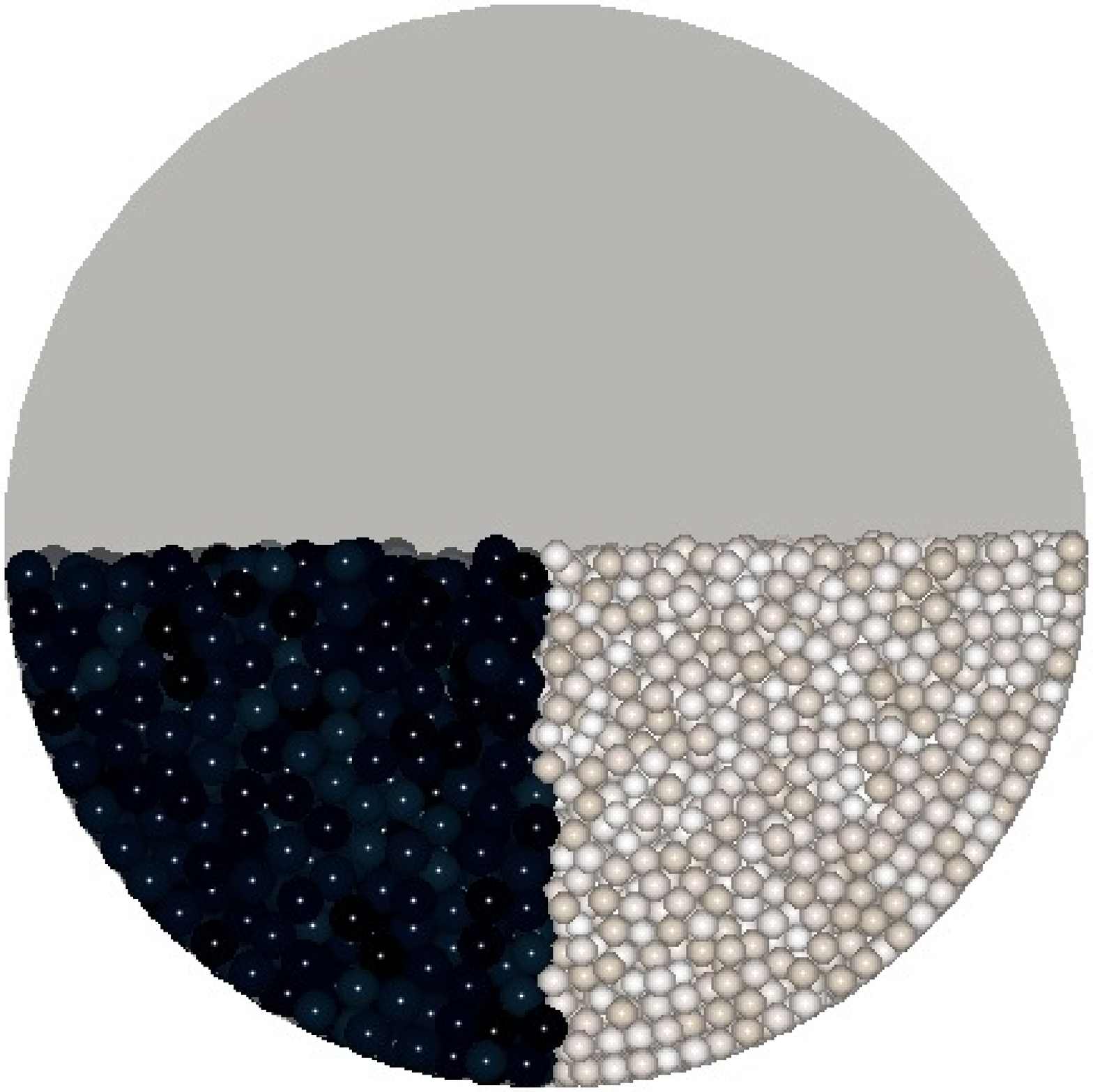,width=0.9\linewidth,angle=-90}\\
\centering \it 0 sec
\end{minipage} 
\begin{minipage}[b]{.49\linewidth}	
\psfig{file=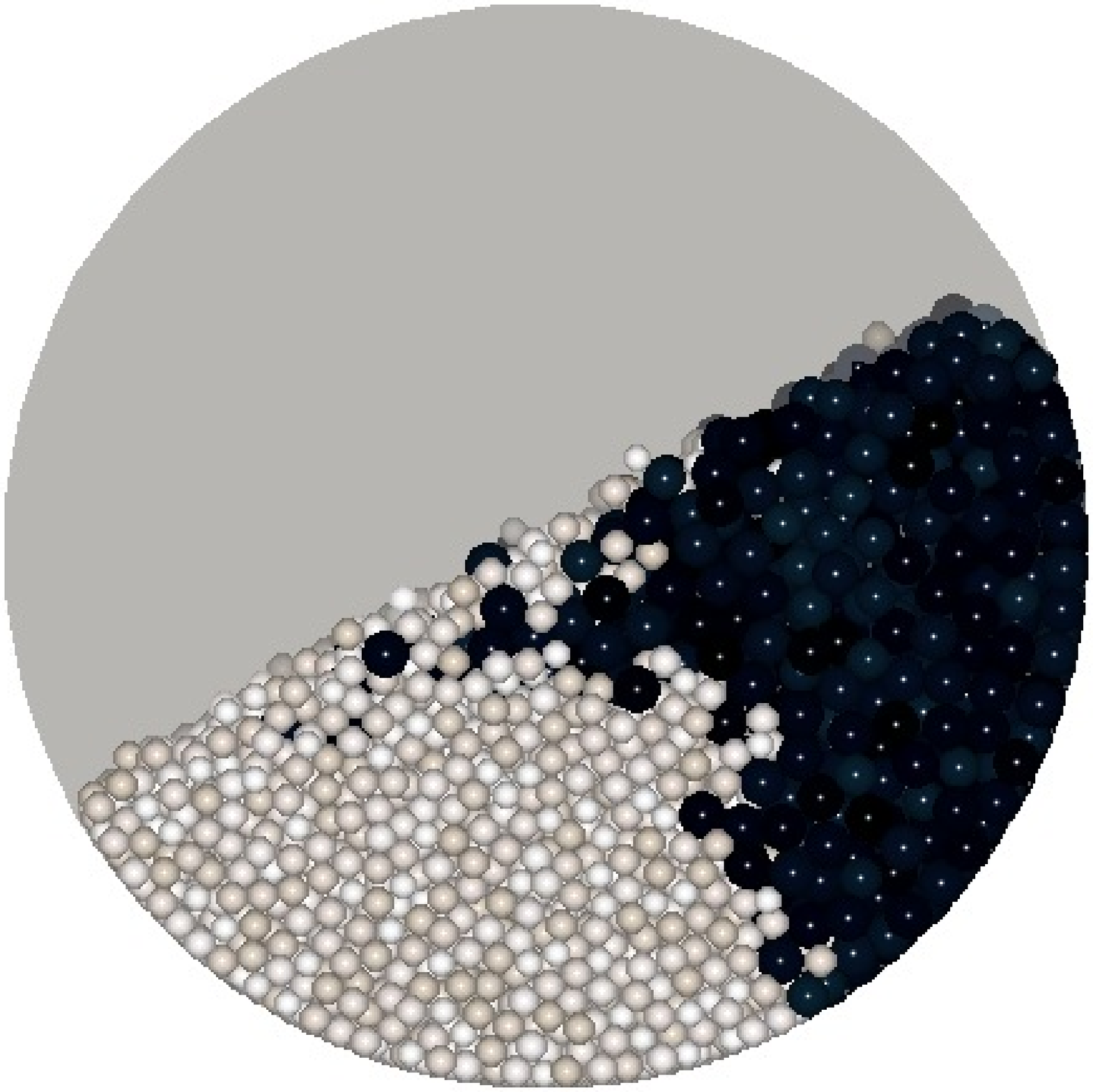,width=0.9\linewidth,angle=-90}\\
\centering \it 1.6 sec
\end{minipage} 
\begin{minipage}[b]{.49\linewidth}	
\psfig{file=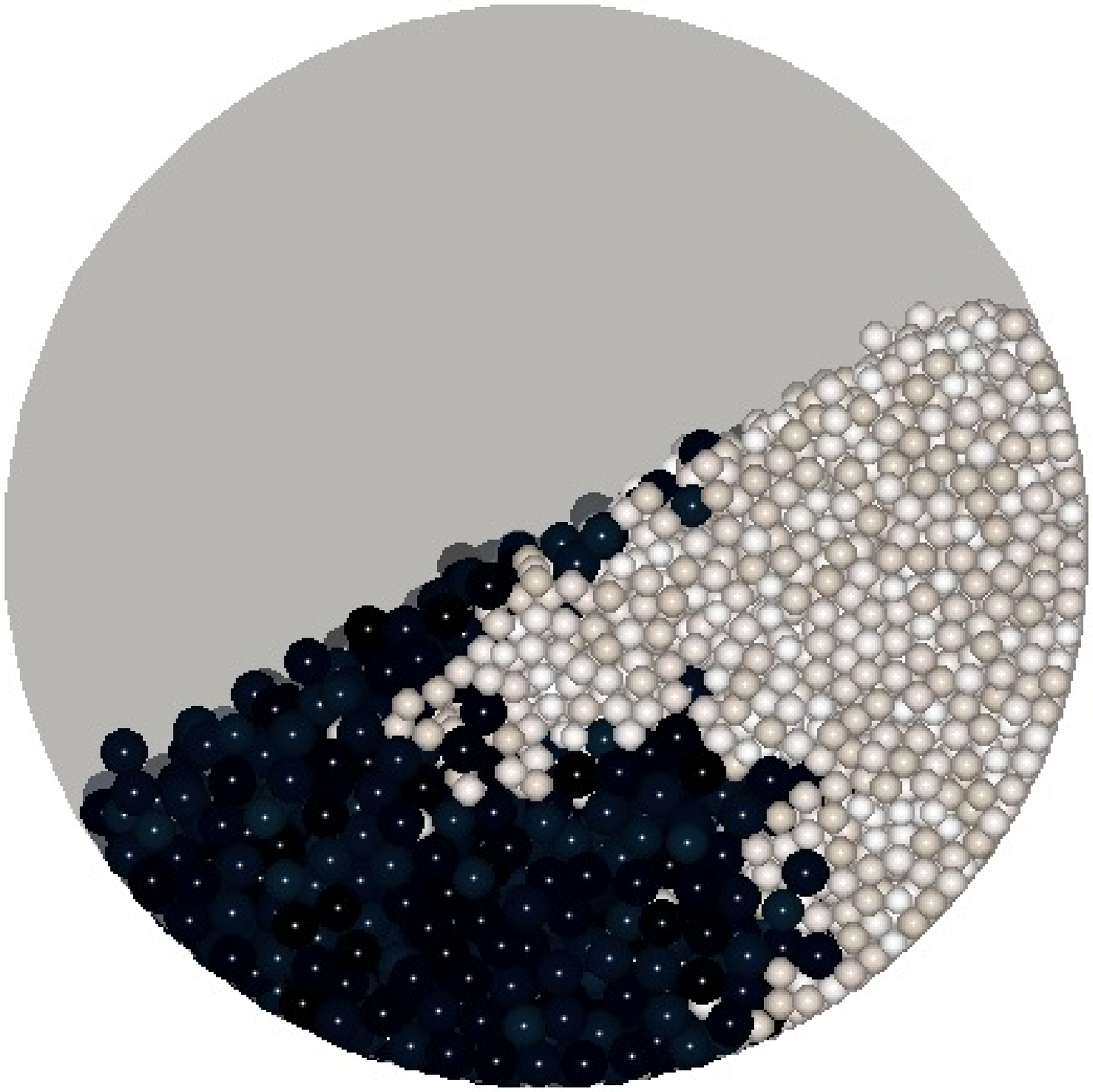,width=0.9\linewidth,angle=-90}\\
\centering \it 2.8 sec
\end{minipage} 
\begin{minipage}[b]{.49\linewidth}	
\psfig{file=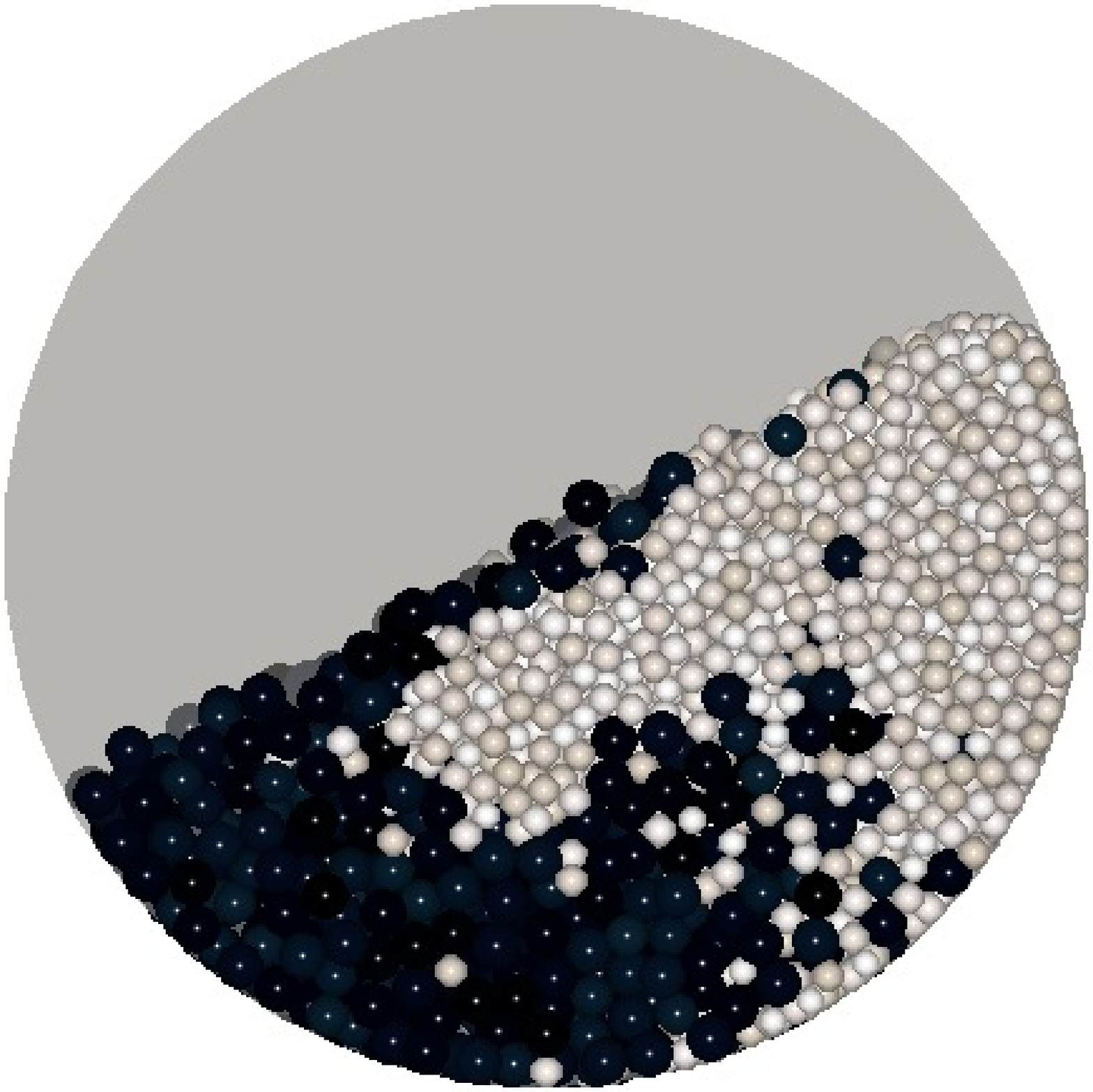,width=0.9\linewidth,angle=-90}\\
\centering \it 5.3 sec
\end{minipage} 
\begin{minipage}[b]{.49\linewidth}	
\psfig{file=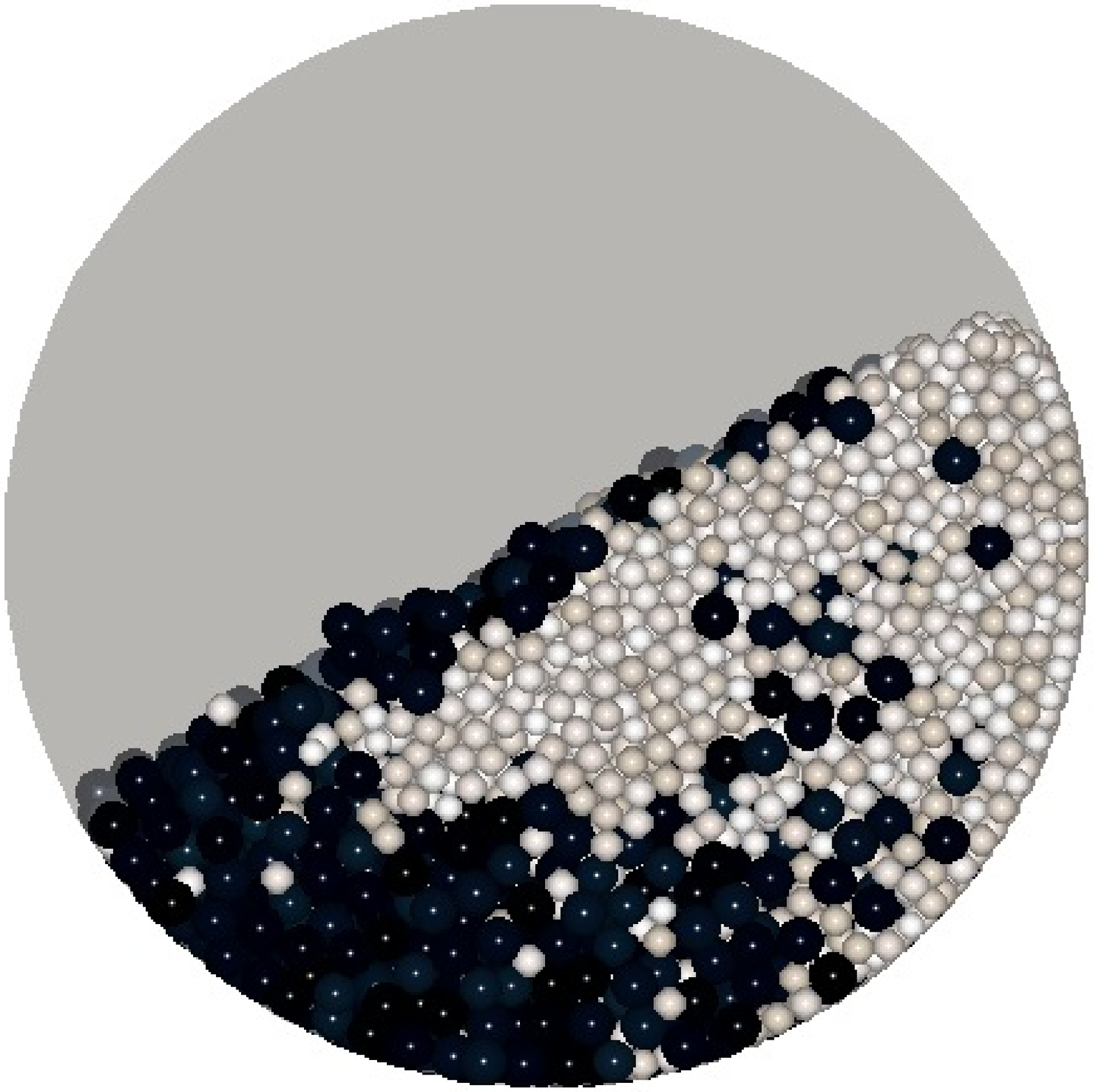,width=0.9\linewidth,angle=-90}\\
\centering \it 7.8 sec
\end{minipage} 
\begin{minipage}[b]{.49\linewidth}	
\psfig{file=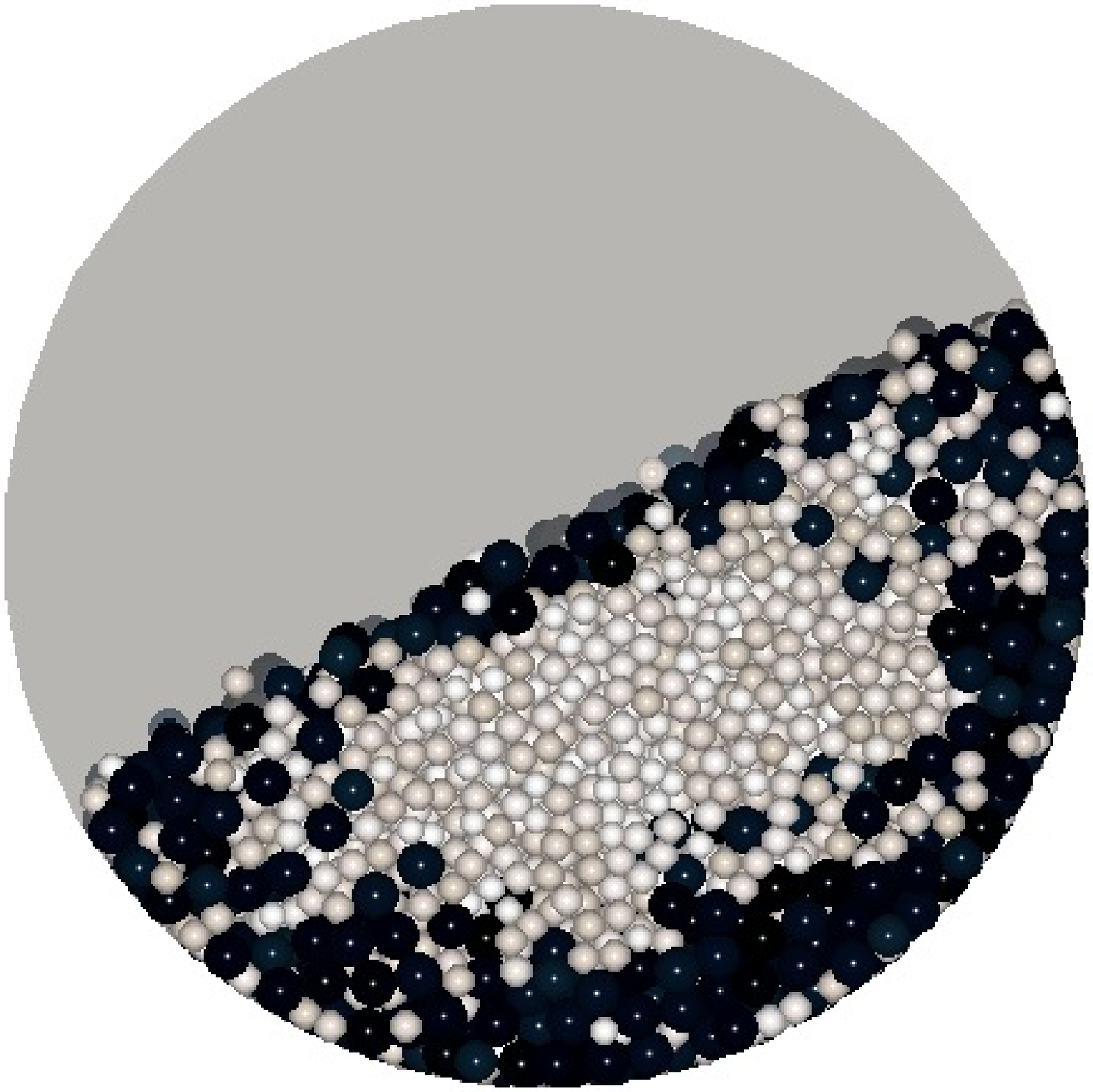,width=0.9\linewidth,angle=-90}\\
\centering \it 56.2 sec
\end{minipage} 
\caption{}
\label{split}
\end{figure}
\newpage
\clearpage

\begin{figure}[htb]
{\centering \input centroid.tex }
 \caption{}
\label{centroid}
\end{figure}
\newpage
\clearpage
\begin{figure}[t]
\noindent	
{\centering \input center.tex }
 \caption{}
\label{center}
\end{figure}
\newpage
\clearpage
\begin{figure}[htb]
\centering
\setlength{\unitlength}{3.0mm}
\begin{picture}(35,40)(-17.5,-22.5)
\Thicklines
\put(0,0){\arc{30}{2.094}{0}}
\qbezier(14.9,0)(14,-12.99)(0,-12.99)
\put(-7.5,-12.99){\line(1,0){8}}
\thinlines
\put(0,0){\arc{30}{0}{2.094}}
\put(-8.84,8.70){\vector(-1,-1){0}}
\put(0,0){\arc{25}{-2.356}{-0.78}}
\put(0,10){\makebox(0,0){$\vec{\omega}$}}
\put(0,0){\line(2,1){13.3}}
\put(6,5){\makebox(0,0){$R$}}
\Thicklines
\put(0,0){\vector(0,-1){6}}
\put(7,-12){\vector(0,1){6}}
\put(7,-12){\vector(-1,0){3.5}}
\put(-18.5,0){\vector(1,0){3.5}}
\put(7,-12){\line(-7,12){3.3}}
\put(3.5,-6){\vector(-2,3){0}}
\thinlines
\put(5,-5.5){\makebox(0,0){$F$}}
\put(-18,2){\makebox(0,0){$F_r$}}
\put(3,-10.5){\makebox(0,0){$F_r$}}
\put(-2,-4){\makebox(0,0){$F_N$}}
\put(9,-6){\makebox(0,0){$F_N$}}
\put(7,-12){\line(-7,12){7}}
\put(7,-12){\line(0,-1){5}}
\put(0,0){\line(0,-1){17}}
\Thicklines
\thinlines
\put(4.5,-16.5){\vector(1,0){2.5}}
\put(2.5,-16.5){\vector(-1,0){2.5}}
\put(3.5,-16.5){\makebox(0,0){$r_0$}}
\thinlines
\put(-15,-17.99){\framebox(30,5)[b]{}}
\end{picture}
\caption{}
\label{sphere}
\end{figure}
\newpage
\clearpage
\setlength{\unitlength}{0.481800pt}
\begin{figure}[htb]
{\centering 
\begin{picture}(1500,850)(0,50)
\tenrm
\thicklines \path(177,179)(197,179)
\thicklines \path(954,179)(934,179)
\put(155,179){\makebox(0,0)[r]{20}}
\thicklines \path(177,275)(197,275)
\thicklines \path(954,275)(934,275)
\put(155,275){\makebox(0,0)[r]{22}}
\thicklines \path(177,371)(197,371)
\thicklines \path(954,371)(934,371)
\put(155,371){\makebox(0,0)[r]{24}}
\thicklines \path(177,466)(197,466)
\thicklines \path(954,466)(934,466)
\put(155,466){\makebox(0,0)[r]{26}}
\thicklines \path(177,562)(197,562)
\thicklines \path(954,562)(934,562)
\put(155,562){\makebox(0,0)[r]{28}}
\thicklines \path(177,658)(197,658)
\thicklines \path(954,658)(934,658)
\put(155,658){\makebox(0,0)[r]{30}}
\thicklines \path(177,754)(197,754)
\thicklines \path(954,754)(934,754)
\put(155,754){\makebox(0,0)[r]{32}}
\thicklines \path(177,849)(197,849)
\thicklines \path(954,849)(934,849)
\put(155,849){\makebox(0,0)[r]{34}}
\thicklines \path(177,945)(197,945)
\thicklines \path(954,945)(934,945)
\put(155,945){\makebox(0,0)[r]{36}}
\thicklines \path(177,179)(177,199)
\thicklines \path(177,945)(177,925)
\put(177,134){\makebox(0,0){0}}
\thicklines \path(307,179)(307,199)
\thicklines \path(307,945)(307,925)
\put(307,134){\makebox(0,0){0.001}}
\thicklines \path(436,179)(436,199)
\thicklines \path(436,945)(436,925)
\put(436,134){\makebox(0,0){0.002}}
\thicklines \path(566,179)(566,199)
\thicklines \path(566,945)(566,925)
\put(566,134){\makebox(0,0){0.003}}
\thicklines \path(695,179)(695,199)
\thicklines \path(695,945)(695,925)
\put(695,134){\makebox(0,0){0.004}}
\thicklines \path(825,179)(825,199)
\thicklines \path(825,945)(825,925)
\put(825,134){\makebox(0,0){0.005}}
\thicklines \path(954,179)(954,199)
\thicklines \path(954,945)(954,925)
\put(954,134){\makebox(0,0){0.006}}
\thicklines \path(177,179)(954,179)(954,945)(177,945)(177,179)
\put(45,562){\makebox(0,0)[l]{\shortstack{$\Theta$}}}
\put(565,67){\makebox(0,0){$r_0[cm]$}}
\put(375,903){\makebox(0,0)[r]{R=1.0mm}}
\Thicklines \path(397,903)(505,903)
\Thicklines \path(397,913)(397,893)
\Thicklines \path(505,913)(505,893)
\Thicklines \path(177,188)(177,198)
\Thicklines \path(167,188)(187,188)
\Thicklines \path(167,198)(187,198)
\Thicklines \path(501,463)(501,472)
\Thicklines \path(491,463)(511,463)
\Thicklines \path(491,472)(511,472)
\Thicklines \path(825,822)(825,842)
\Thicklines \path(815,822)(835,822)
\Thicklines \path(815,842)(835,842)
\Thicklines \path(177,344)(177,363)
\Thicklines \path(167,344)(187,344)
\Thicklines \path(167,363)(187,363)
\Thicklines \path(501,603)(501,612)
\Thicklines \path(491,603)(511,603)
\Thicklines \path(491,612)(511,612)
\Thicklines \path(825,868)(825,888)
\Thicklines \path(815,868)(835,868)
\Thicklines \path(815,888)(835,888)
\Thicklines \path(177,442)(177,461)
\Thicklines \path(167,442)(187,442)
\Thicklines \path(167,461)(187,461)
\Thicklines \path(501,688)(501,697)
\Thicklines \path(491,688)(511,688)
\Thicklines \path(491,697)(511,697)
\Thicklines \path(825,925)(825,944)
\Thicklines \path(815,925)(835,925)
\Thicklines \path(815,944)(835,944)
\put(177,193){\raisebox{-1.2pt}{\makebox(0,0){$\Box$}}}
\put(501,468){\raisebox{-1.2pt}{\makebox(0,0){$\Box$}}}
\put(825,832){\raisebox{-1.2pt}{\makebox(0,0){$\Box$}}}
\put(177,354){\raisebox{-1.2pt}{\makebox(0,0){$\Box$}}}
\put(501,607){\raisebox{-1.2pt}{\makebox(0,0){$\Box$}}}
\put(825,878){\raisebox{-1.2pt}{\makebox(0,0){$\Box$}}}
\put(177,452){\raisebox{-1.2pt}{\makebox(0,0){$\Box$}}}
\put(501,693){\raisebox{-1.2pt}{\makebox(0,0){$\Box$}}}
\put(825,934){\raisebox{-1.2pt}{\makebox(0,0){$\Box$}}}
\put(451,903){\raisebox{-1.2pt}{\makebox(0,0){$\Box$}}}
\Thicklines \path(177,193)(177,193)(501,468)(825,832)
\Thicklines \path(177,354)(177,354)(501,607)(825,878)
\Thicklines \path(177,452)(177,452)(501,693)(825,934)
\put(375,858){\makebox(0,0)[r]{R=1.5mm}}
\thinlines \path(397,858)(505,858)
\thinlines \path(397,868)(397,848)
\thinlines \path(505,868)(505,848)
\thinlines \path(177,282)(177,292)
\thinlines \path(167,282)(187,282)
\thinlines \path(167,292)(187,292)
\thinlines \path(501,423)(501,433)
\thinlines \path(491,423)(511,423)
\thinlines \path(491,433)(511,433)
\thinlines \path(825,569)(825,579)
\thinlines \path(815,569)(835,569)
\thinlines \path(815,579)(835,579)
\thinlines \path(177,461)(177,471)
\thinlines \path(167,461)(187,461)
\thinlines \path(167,471)(187,471)
\thinlines \path(501,591)(501,610)
\thinlines \path(491,591)(511,591)
\thinlines \path(491,610)(511,610)
\thinlines \path(825,725)(825,744)
\thinlines \path(815,725)(835,725)
\thinlines \path(815,744)(835,744)
\thinlines \path(177,572)(177,581)
\thinlines \path(167,572)(187,572)
\thinlines \path(167,581)(187,581)
\thinlines \path(501,672)(501,701)
\thinlines \path(491,672)(511,672)
\thinlines \path(491,701)(511,701)
\thinlines \path(825,816)(825,844)
\thinlines \path(815,816)(835,816)
\thinlines \path(815,844)(835,844)
\put(177,287){\raisebox{-1.2pt}{\makebox(0,0){$\Diamond$}}}
\put(501,428){\raisebox{-1.2pt}{\makebox(0,0){$\Diamond$}}}
\put(825,574){\raisebox{-1.2pt}{\makebox(0,0){$\Diamond$}}}
\put(177,466){\raisebox{-1.2pt}{\makebox(0,0){$\Diamond$}}}
\put(501,600){\raisebox{-1.2pt}{\makebox(0,0){$\Diamond$}}}
\put(825,734){\raisebox{-1.2pt}{\makebox(0,0){$\Diamond$}}}
\put(177,576){\raisebox{-1.2pt}{\makebox(0,0){$\Diamond$}}}
\put(501,686){\raisebox{-1.2pt}{\makebox(0,0){$\Diamond$}}}
\put(825,830){\raisebox{-1.2pt}{\makebox(0,0){$\Diamond$}}}
\put(451,858){\raisebox{-1.2pt}{\makebox(0,0){$\Diamond$}}}
\thinlines \path(177,287)(177,287)(501,428)(825,574)
\thinlines \path(177,466)(177,466)(501,600)(825,734)
\thinlines \path(177,576)(177,576)(501,686)(825,830)
\end{picture} }
 \caption{}
\label{theta_rolling}
\end{figure}
\fi
\end{document}